\documentclass[a4paper,fleqn,usenatbib]{mnras}
\usepackage{graphicx}
\usepackage{graphics}
\usepackage{amsmath}
\usepackage{multirow}
\usepackage{color}
\usepackage{xcolor}
\usepackage{bm}		

\usepackage{float}

\def\kmskpc{{\rm\,km\, \,s^{-1} \, {kpc}^{-1}}}

\def\deg{{^\circ}}

\def\mathnew{\mathsurround=0pt}   
\def\simov#1#2{\lower .5pt\vbox{\baselineskip0pt  
    \lineskip-.5pt\ialign{$\mathnew#1\hfil##\hfil$\crcr#2\crcr\sim\crcr}}}

\def\'#1{\ifx#1i{\accent"13\i}\else{\accent"13#1}\fi}

\def\aap{{A\&A}}

\def\aj{{AJ}}
\def\apj{{ApJ}}
\def\apjl{{ApJL}}

\def\mnras{{MNRAS}}
\def\apjs{{ApJS}}

\def\rmxaa{{Rev. Mex. Astron. Astrofis.}}

  \title[MW Spiral Arms from the MDF and Radial Mixing] {Dynamics and Morphology of the Milky Way Spiral
    Arms from the Metallicity Distribution and Radial Mixing}
\author[Martinez-Medina et al. 2016]{L.A. Martinez-Medina
  \thanks{Contact e-mail:
    \href{mailto:lamartinez@astro.unam.mx}{lamartinez@astro.unam.mx}},
  B. Pichardo,  A. Peimbert, \& L. Carigi \\ Instituto de Astronom\'ia,
  Universidad Nacional Aut\'onoma de M\'exico, A.P. 70--264, 04510,
  M\'exico, CDMX, M\'exico}

\begin{document}
\label{firstpage}
\pagerange{\pageref{firstpage}--\pageref{lastpage}}
\maketitle

\begin{abstract}
Albeit radial migration must be a ubiquitous process in disc galaxies,
its significance in the evolution of stellar discs is not always
reflected through global trends. However, there are other key
observables, such as the metallicity distribution function (MDF), that
may shed some light in this matter. We argue that the shape of the MDF
not only tells us whether the stellar disc experienced radial
migration, but it also contains important clues on the structure that
triggered it. Specifically, the MDF contains information about the
dynamics and morphology of the spiral pattern. To constrain the spiral
parameters, we have included a detailed chemical tagging in our
simulations; this allows us to produce a restriction of the structural
parameters of the spiral arms in the Milky Way as well as a method to
constrain chemical evolution models towards the center of the Galactic
disc, where no chemical model provides information. We also found that
it is unlikely that the Sun was formed near its current galactocentric
position, therefore it might be inaccurate to consider the Sun as
representative of the chemical abundances in the solar
neighborhood. We also show that a stellar disc of the Milky Way, after
evolving dynamically and chemically for 5 Gyr, preserves 80\% of its
original global metallicity gradient despite having suffered important
heating and radial migration; this means that the presence of a
metallicity gradient in a given galaxy, does not guarantee that radial
mixing has not played a role in its evolution.

\end{abstract}                
 
\begin{keywords}
Galaxy: disc --- Galaxy: evolution --- Galaxy: kinematics and dynamics --- Galaxy: structure --- Galaxy: abundances --- Sun: abundances
\end{keywords}


\section{Introduction} \label{sec:intro}

It has been a long journey since the first clues of the discrepancy
between our ideal picture, of a well behaved axisymmetric Galactic
disc, and observations were revealed. In this ideal scenario stellar
orbits were approximately circular, the large scale non-axisymmetric
structures (spiral arms and bar) had negligible effects on the
dynamics, and a perfect correlation between age and metallicity of a
star could be found for a given Galactic region. Some of the
pioneering work noticing the discrepancy between the required
characteristics for that ideal scenario and reality, comes with
\citet{1972ade..coll...55G,1989Ap&SS.156...29G}. He found in the solar
neighbourhood a super metal-rich stellar population that exhibits the
abundance of chemical elements and kinematics of the central regions
of the Galaxy; since that time a possible stellar mixing that somehow
was taking place in the Galaxy was proposed
\citep[e.g.][]{1977A&A....60..263W}. Likewise,
\citet{1993A&A...275..101E} found a poor correlation between age and
metallicity for stars in the Solar neighbourhood from the thin
disc. Later on \citet{1993A&A...280..136F} produced the first analysis
of the relevance of radial mixing on the chemical properties of the
Galaxy; by using the spread observed in the age metallicity relation
and in the [$\alpha$/Fe] $vs.$ [Fe/H] relation for stars in the solar
neighbourhood, the authors concluded that the observed spread can be
explained by stars coming from different regions of the Galaxy.

Since this effect is not readily explained by plain orbital excursions
on epicycles around their birth place \citep[hereafter
  SB02]{Sellwood2002}, i.e. as an effect of simple heating by the
large-scale Galactic structures (arms and bar), another mechanism was
invoked: stellar migration (also known as churning). In this paper we
will use the definition of SB02; for them, stellar migration is the
redistribution of angular momentum for stars produced mainly by their
interaction with the large scale non-axisymmetric structures of a
galaxy. This process, takes place mainly in the corotation resonance
region of a given pattern and has the important characteristic of
changing angular momentum of stellar orbits while keeping their
orbital eccentricity unchanged.

In \citet{MartinezMedina2016}, we studied extensively the change of
angular momentum induced by the large scale structures of the
Galaxy. We found that the bar produces radial migration mainly in the
region between 2 and 6 kpc, while the spiral arms dominate for $R>6$
kpc. This means that, while in the inner region of the disc the
migration is dominated by the joint action of the bar and spiral arms,
at outer radii, radial migration is mainly dominated by the spiral
arms. We show that the spiral arms are able to imprint their signature
in the radial migration with a minimal participation of the bar.

Radial migration (or churning), unlike simple dynamical heating (also
known as blurring) that increases the orbital eccentricity, results
untrackable since it is not expected to leave kinematic imprints on
the spatial origin of the orbits. This would pose a challenge for
chemical models because, with increasing flattening due to radial
migration, the quantification of the past metallicity gradients (of
the stellar population of galaxies) would become increasingly
inaccurate.

A lot of work has been devoted to the phenomenon of radial mixing
since it was recognized in the 70´s. The most recent papers on the
subject refer to bars and/or spiral arms as the main mechanisms
inducing radial mixing. In the case of the spiral arms, stars close to
the corotation region, particularly those with high angular momentum
orbits, will suffer radial migration more efficiently due to resonant
scatter
\citep[e.g. SB02;][]{2008ApJ...684L..79R,2012MNRAS.426.2089R,Schonrich2009,2009MNRAS.399.1145S,VeraCiro2014}. Some
of the proposals for the mechanisms to produce efficient radial
migration (as well as radial displacements) are: fixed corotation long
lived spiral arms, usually represented by simple analytical models
\citep[e.g.][]{2003ApJ...589..210L,2009IAUS..254..381P}; transient
recurrent spiral arms that change the corotation resonance along large
radial extensions of the discs
\citep[]{Schonrich2009,2012MNRAS.421.1529G,2012MNRAS.426..167G,2011ApJ...737....8L};
bar and spiral arms resonance overlapping that induces a more
efficient redistribution of angular momentum in the disc
\citep[]{Minchev2010,2011ApJ...733...39S,2011A&A...534A..75B};
short-lived transient density peaks originated by interfering spiral
patterns \citep[]{2012arXiv1207.5753C}, and short-lived, recurrent
grand design spirals \citep[]{2012MNRAS.426L..46A}.

Although the effect of stellar migration on chemical evolution in
galaxies is still not quite clear, it is undeniable its impact on the
shape of the Milky Way (MW) MDF
\citep{2011A&A...530A.138C,Hayden2015,2016ApJ...818L...6L,MartinezMedina2016}. In
particular \citet{2013MNRAS.436.1479K} stress the relevance of three
factors that contribute to the relation between radial migration and
chemical evolution: the strength of the bar or spiral arms (the
stronger the structures the larger the effects of radial migration and
the larger the effects on the chemical evolution); the duration of the
radial migration (the longer, the larger the effects); and the slope
of the metallicity curves for a given galaxy (the steeper, the larger
the effects). Based on the results of the present work, we concur on
the fact that the relation between dynamics (radial migration) and
chemistry, depends on the particular galaxy under study.

In this work we have produced a comprehensive orbital study for the MW
disc; we employed a novel recipe of chemical tagging as well as an
observationally motivated potential model for the MW, that includes
the bar and a very detailed three dimensional density based spiral
arms potential. Our study has three purposes, the first one is to
produce restrictions for the structural and dynamical parameters of
both the bar and the spiral arms. The second, to provide a restriction
to the abundance gradient of the disc (at all times and radii), all
based on an observable in our Galaxy: the MDF. Lastly, by considering
plausible Galactic parameters for the MW, we show that the chemical
gradient is nearly conserved at all times in the history of the
Galaxy, i.e., neither radial migration nor heating remove the large
scale chemical gradient of the MW Galactic disc.

This paper is organised as follows. The galactic model, initial
conditions, and methodology are described in Section \ref{model}. A
study on radial migration, radial heating and the theoretical MDF is
presented on Section \ref{Results}. In Section \ref{tagging}, we
present the chemical curves and tagging recipe employed in this work.
Our best fit for the dynamical and structural parameters to the MW MDF
is shown in Section \ref{shape}. Restrictions to the chemical gradient
in the inner and outer regions of the MW disc are in Section
\ref{constrain}. In Section \ref{stellardynamics}, we show the studies
of the dynamic effects of the spiral arms and bar imprinted on the
MDF. We show the radial reach of migration and heating in Section
\ref{radialreach}. In Section \ref{MAPs}, we present the resulting MDF
for populations of different ages. Finally, we present the discussion
and conclusions, in Sections \ref{discussion} and \ref{conclusions},
respectively.

\section{The Galaxy Model}
\label{model}
As already described in depth in \citet{MartinezMedina2016}, for this
work we have employed an elaborate, three dimensional (although
stationary) model, adjusted to the best of recent knowledge of the MW
Galaxy structural and dynamical parameters instead of employing the
more sophisticated N-body simulations. N-body simulations are not
applicable to achieving the goals of this work for the following
reasons: a) our model is completely adjustable (contrary to N-body
simulations); b) it is considerably faster, adequate to statistical
studies like the ones we are presenting in this work; c) because of
its nature, this type of models allows us to study in great detail
individual stellar orbital behaviour without the known resolution
problems of N-body simulations.

The model is composed by an axisymmetric part and a non-axisymmetric
part (spiral arms and bar). The axisymmetric potential is based on the
potential of \citet{AS91}, that consists of a Miyamoto-Nagai single
disc with a vertical height of 250 pc, a spherical bulge, also
Miyamoto-Nagai, and a supermassive spherical halo. The spiral arms and
the bar are introduced adiabatically by reducing the mass of the disc
and the bulge respectively. The Galactic potential is scaled to the
Sun's galactocentric distance, $R_0$ = 8.5 kpc, and the local rotation
velocity, ${\Theta}_0$ = 220 km s$^{-1}$.

For the spiral arms we employ a model formed by a two-armed 3D density
distribution made of individual inhomogeneous oblate spheroids, that
we call the PERLAS model \citep{PMME03}. The density is distributed as
an exponential decline along the arms. The spheroids act as bricks to
construct the spiral arms structure; they are located along a
logarithmic spiral locus. PERLAS is completely adjustable (i.e. the
width, height, scale lengths, density fall along the spiral arms and
transversal to them, etc.) to better represent the available
observations of the spiral arms in the Milky Way. The total mass of
the spiral arms taken in these experiments is $4.28 \times 10^9$
M$_{\sun}$, that corresponds to a mass ratio of Marms/Mdisc = 0.05.

For the bar we have selected a triaxial inhomogeneous ellipsoid built
as a the superposition of a large number of homogeneous ellipsoids to
achieve a smooth density fall \citep{Pichardo2004}. The model
approximates the density fall fitted by \citet{1998ApJ...492..495F}
from the COBE/DIRBE observations of the Galactic centre. The total
mass of the bar is $1.4 × 10^{10}$ M$_{\sun}$, within the
observational limits. Regarding the angular speed, a long list of
studies have estimated this parameter concluding that the most likely
value lies in the range $\Omega = 45 - 60$ km s$^{-1}$ kpc$^{-1}$. We
adopt the value $\Omega = 45$ km s$^{-1}$ kpc$^{-1}$, based on
the formation of moving groups in the solar neighbourhood.

For all details on the construction and the fit to observations see
\citet{2015ApJ...802..109M,MartinezMedina2016,PMME03,Pichardo2004,2012AJ....143...73P,2015MNRAS.451..705M};
in particular for the present work we have adopted the parameters
presented in Section 2 of \citet{MartinezMedina2016}. Brief
descriptions of the parameters employed in each simulation are
presented across the paper.

\section{Radial migration, radial heating, and the shape of the MDF}
\label{Results}

In a disc galaxy, stars orbiting the galactic center exchange angular
momentum with non-axisymmetric substructure in the disc. The
redistribution of angular momentum in the stellar orbits produces
radial migration as well as radial heating, both displace stars from
its current galactocentric radius and together encompass what is
called radial mixing. 

Even when radial migration does not leave a kinematical imprint in the
stellar disc, the mechanism can impact some other observables because
it can radially displace stars by several kiloparsecs. 

Recently there has been found evidence of radial migration in the MW
\citep{Hayden2015}, confirming its ability of shaping the MDF
\citep{2016ApJ...818L...6L, MartinezMedina2016}.

In \citet{MartinezMedina2016}, by comparing three galactic models: one
with a central bar, one with spiral arms, and one with bar + spiral
arms, we found that the change in the skewness of the MDF is mainly
driven by the spiral pattern.  This means that, through radial
migration, the spirals left an imprint in the metallicity distribution
of the stellar disc.

\subsection{The birth radius as a proxy for metallicity}
\label{Proxy}

Since a star preserves information of the state of the ISM at its
birth place and epoch, there is a correlation between the birth radius
of a star and its metallicity.  In order to study the extent of the
stellar radial displacements due to radial migration, in
\citet{MartinezMedina2016}, we used the birth radius of a star as a
proxy for its metallicity. As expected, the initial radii
distributions exhibit a change in skewness when going from the inner
to the outer disc. Moreover, in the outer disc, we found that the
spiral arms are the main responsible for shaping the initial radii
distributions and hence, the MDFs.

\begin{figure}
\begin{center}
\includegraphics[width=9cm]{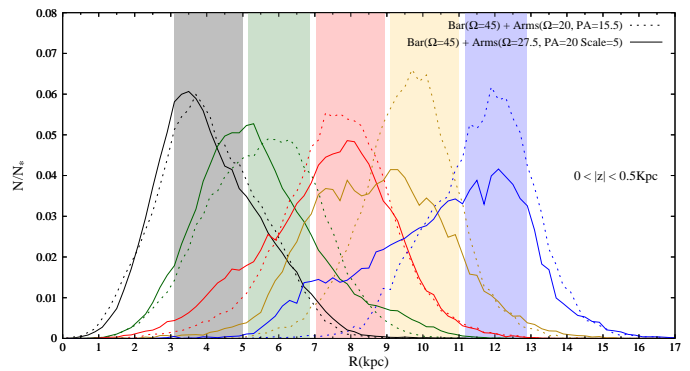}
\end{center}
\caption{Initial galactocentric radii distribution of stars that at
  the end of the simulation are located within one of the five
  coloured bins. Each curve corresponds to the closest same-colour
  shaded bin. $N$ is the number of stars at a given initial radius
  and $N_*$ is the total number of stars that have ended within each
  coloured bin.}
\label{f1}
\end{figure}

By using the proxy of the birth radius for metallicity, we show how
the shape of the MDF depends on the structural parameters of the
spiral arms. In Fig \ref{f1} we plot the initial radii
distribution for two MW models; as indicated in the Figure, the models
differ in the structural parameters of the spiral pattern; here we can
see the sensibility of the shape of the curves with the amplitude of
the spirals and the location of the corotation radius. The purpose of
comparing these two models is to show that the initial radii
distributions (and hence the MDF) contains information about the
structural parameters of the spiral arms.

Although this proxy provides us insight about how the MDF would look
like, in the next sections we complement our simulations with a
chemical tagging in order to analyse and make a more systematic
comparison of the simulated MDFs with those observed in the MW.

\section{Chemical tagging}
\label{tagging}

\subsection{The Chemical Evolution Model}
The evolution of [Fe/H] and Star Formation Rate used in our
simulations is obtained from a Chemical Evolution Model for the Galaxy
by \citet{2011RMxAA..47..139C}. This model was built to match three
observational constraints along the Galactic disc: the radial
distributions of the surface density of both the total baryonic mass
as well as the gas mass, and the O/H radial gradient from \ion{H}{II}
regions located between 6 and 11 kpc. The O/H gaseous values,
determined from recombination lines by
\citet[][]{2007ApJ...670..457G}, were increased due to the correction
by dust depletion \citep{2010ApJ...724..791P}.

The main assumptions of the chemical model are: a) the MW disc was
formed in an inside-out scenario from primordial accretion with
time-scales $\tau(R) = ((R/$kpc$)-2)$ Gyr, where $R$ is the
galactocentric distance; b) for each $R$, a double-accretion scenario
was assumed, that is, the halo was built with a timescale of $\tau'(R)
= 0.5$ Gyr during the first Gyr and then the disc formed with
$\tau(R)$, till 13 Gyr (the present day); c) the star formation rate
was similar to the Kennicutt law \citep{1998ApJ...498..541K}, and its
efficiency was five times higher during the halo-forming phase than
during the disc-forming phase; d) the initial mass function of
\citet{1993MNRAS.262..545K} was applied in the $0.08-80$ M$_\odot$
range; e) an array of metal-dependent stellar yields for He, C, N, O,
Fe, and, $Z$ (all heavy elements) has been used, characterized mainly
by stellar yields for massive stars that consider an intermediate
mass-loss rate due to stellar winds \citep[see IWY model
  by][]{2011RMxAA..47..139C}; f) when modeling the chemical evolution
of the gas, neither radial flows of gas nor radial migration of stars
were considered (note that although the dynamical model feeds of the
chemical model, the chemical model does not feed of the dynamical
one); g) the chemical evolution model was computed for $R\ge3$ kpc,
because the Galactic bulge is located in the region $R \la 3$ kpc and
its evolution is quite different to that of the Galactic disc.

As by-products, the model reproduces very well: a) the radial
distribution of the stellar mass and the star formation rate; b) the
C/H, N/H, O/H, N/O, C/O, and Fe/H gradients; and c) the C/Fe vs Fe/H,
O/Fe vs Fe/H, and C/O vs O/H trends shown by stars of different ages
in the solar vicinity ($R\approx8.5$ kpc)
\citep{2011RMxAA..47..139C,2013MNRAS.433..382E,2016ApJ...827..126B}.

\begin{figure}
\begin{center}
\includegraphics[width=9cm]{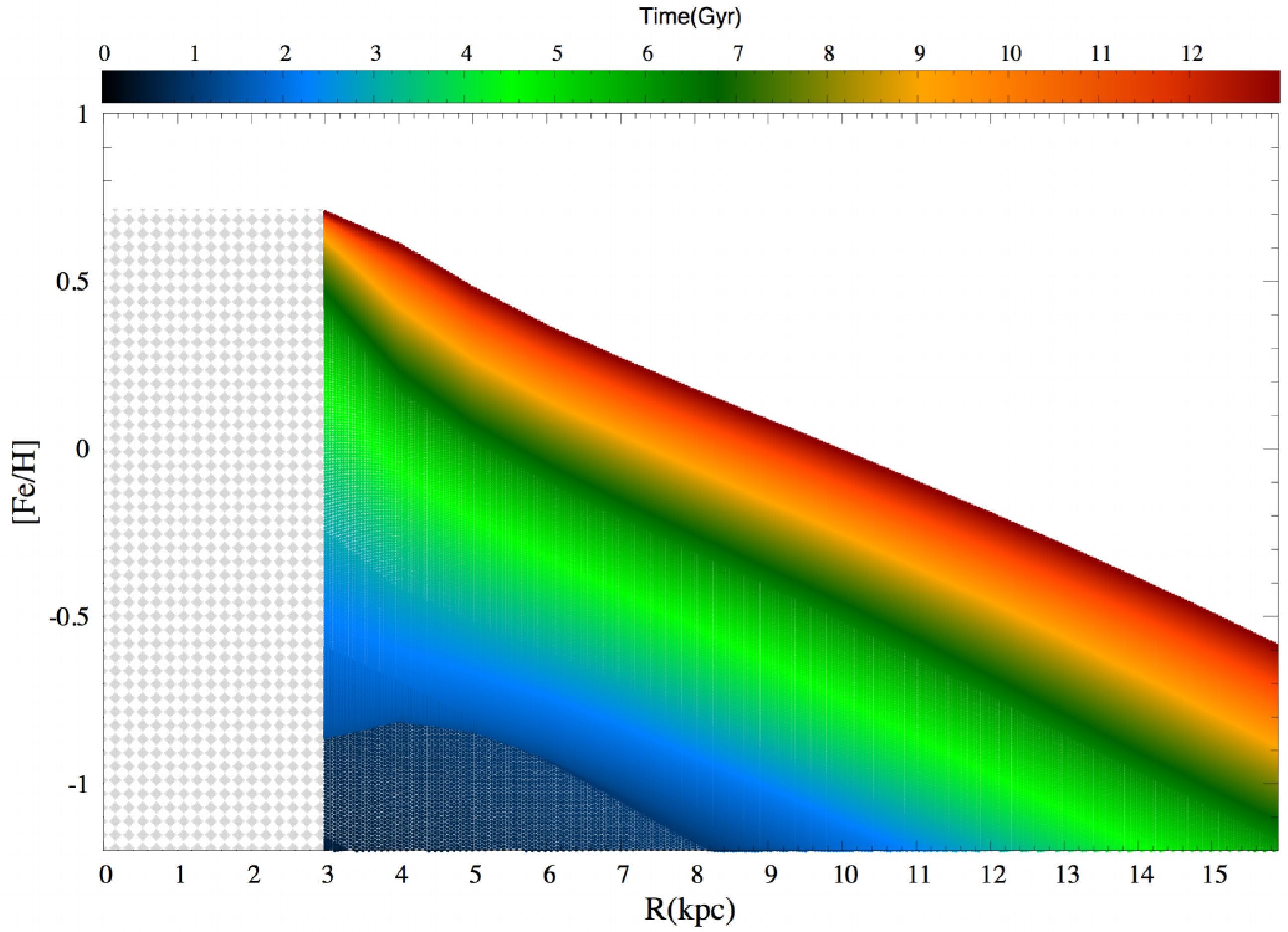}
\end{center}
\caption{[Fe/H] curves for the chemical evolution model of the MW. The
  color code traces the history of the Galactic disc, from its
  formation to the present time (13 Gyr). }
\label{f2}
\end{figure}

In Figure \ref{f2} we present the original output from the
chemical model. The data are shown in an array of metallicity values
for galactocentric distances between $3$ $\le$ $R$ $\le$ $20$ kpc, going
in time from 0 to 13 Gyr (present time), with a high temporal
resolution.

\begin{figure}
\begin{center}
\includegraphics[width=9cm]{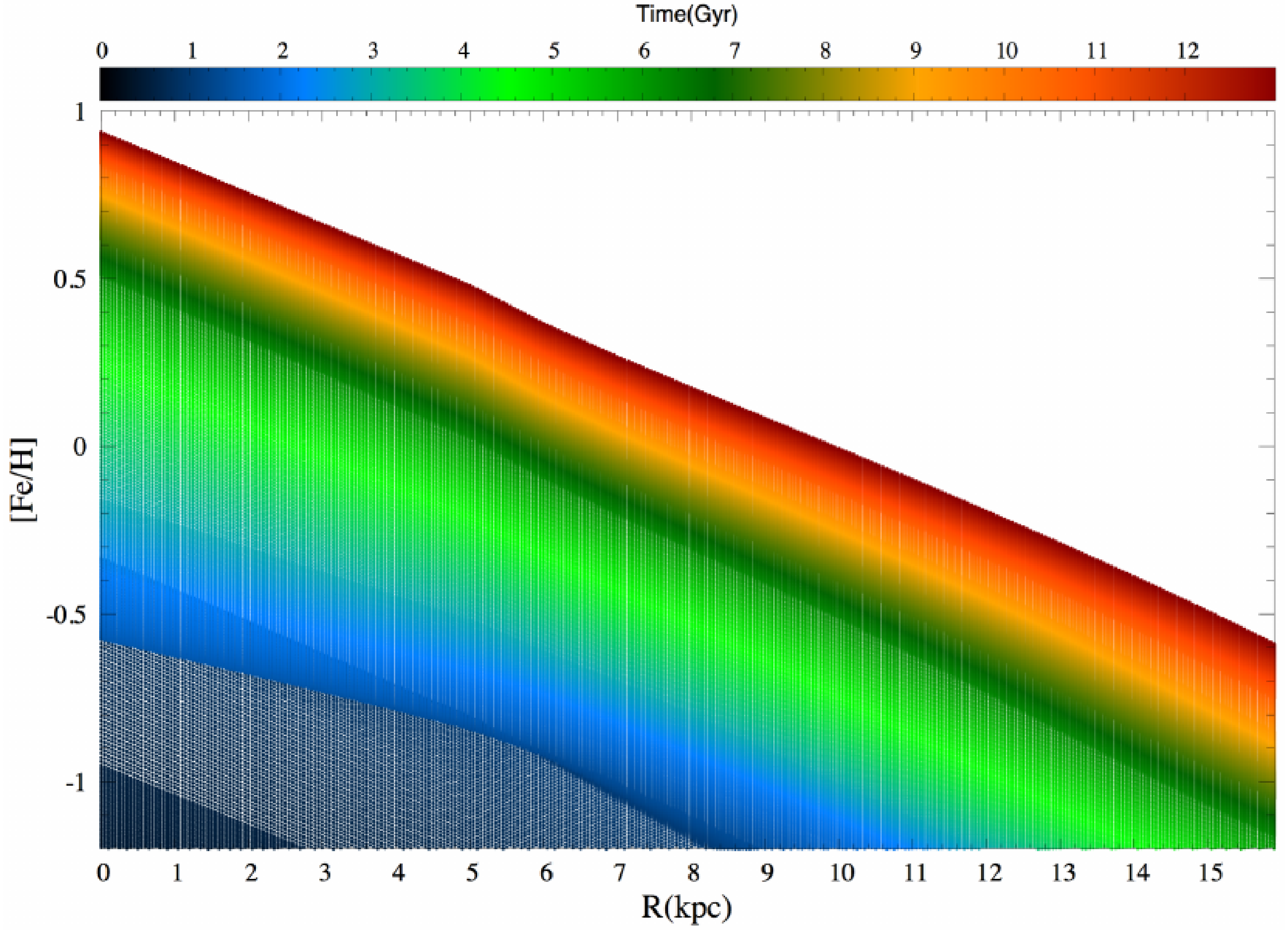}
\end{center}
\caption{[Fe/H] curves extrapolated towards the inner 3 kpc. The color
  code traces the history of the Galactic disc, from its formation to
  the present time (13 Gyr).}
\label{f3}
\end{figure}

As mentioned before, it is uncommon that chemical evolution models
provide information of the inner regions of the disc, this due to the
lack of observational constraints as well as difficulties in modeling
the gaseous component of the Galactic bar and bulge. Such
incompleteness leads us to take an assumption for the chemical content
of the inner disc, $R$ $<$ $3$ kpc, as shown in Figure \ref{f3} where
an extrapolation towards the center provides us with a complete data
set. Let us anticipate that the extrapolation is not arbitrary, it is
motivated by the MDFs themselves, as further explained in Section
\ref{constrain}. Also, while this is our initial hypothesis for the
chemical gradient, the observed MDFs require a slightly flatter
gradient, which we also explain in Section \ref{constrain}.

\begin{figure}
\begin{center}
\includegraphics[width=9cm]{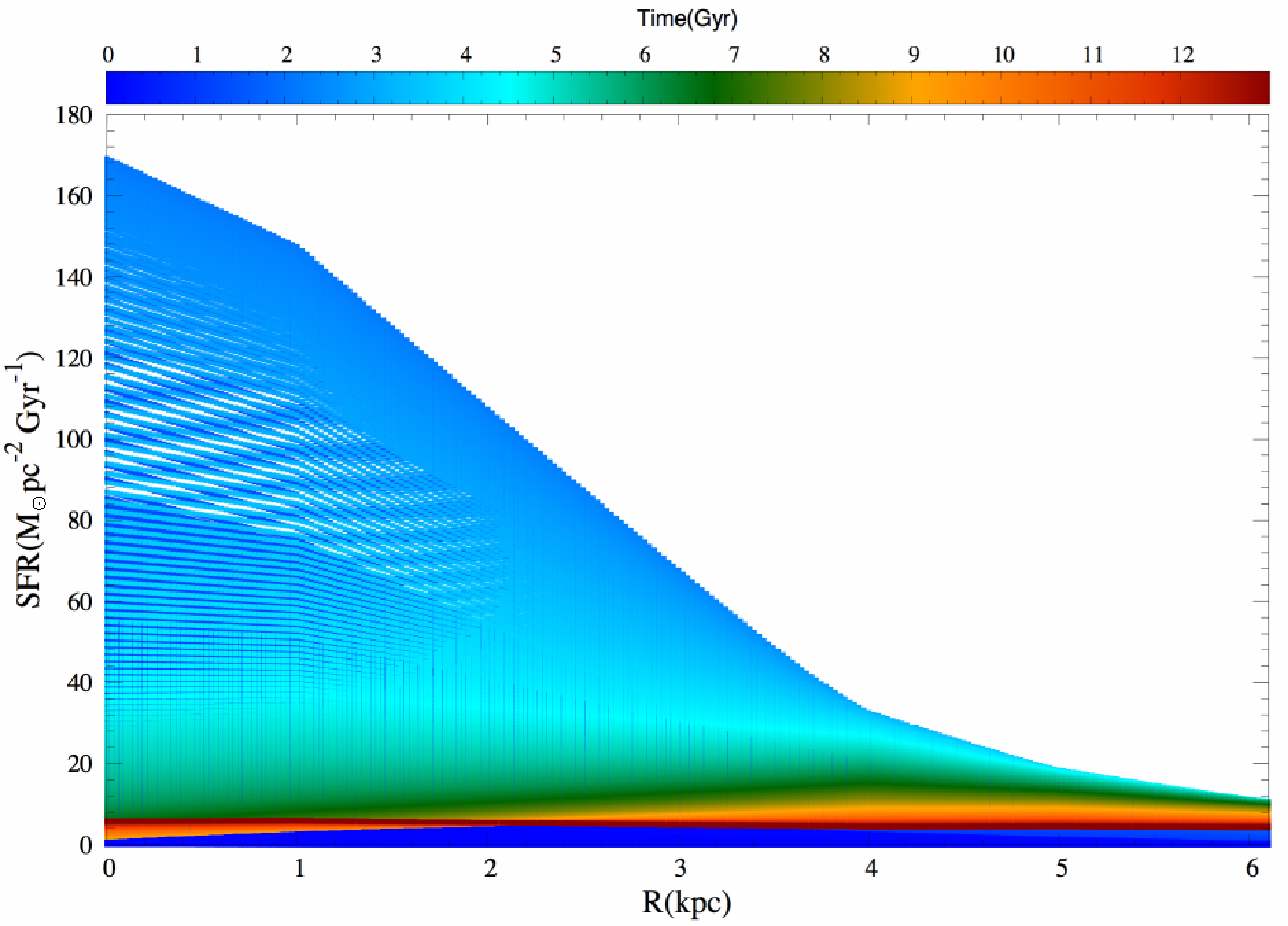}
\end{center}
\caption{SFR in the central region of the disc as a function of
  galactocentric radius and time, including an extrapolation towards
  the inner 3 kpc.  The color code traces the history of the Galactic
  disc, from its formation to the present time (13 Gyr).}
\label{f4}
\end{figure}

\begin{figure}
\begin{center}
\includegraphics[width=9cm]{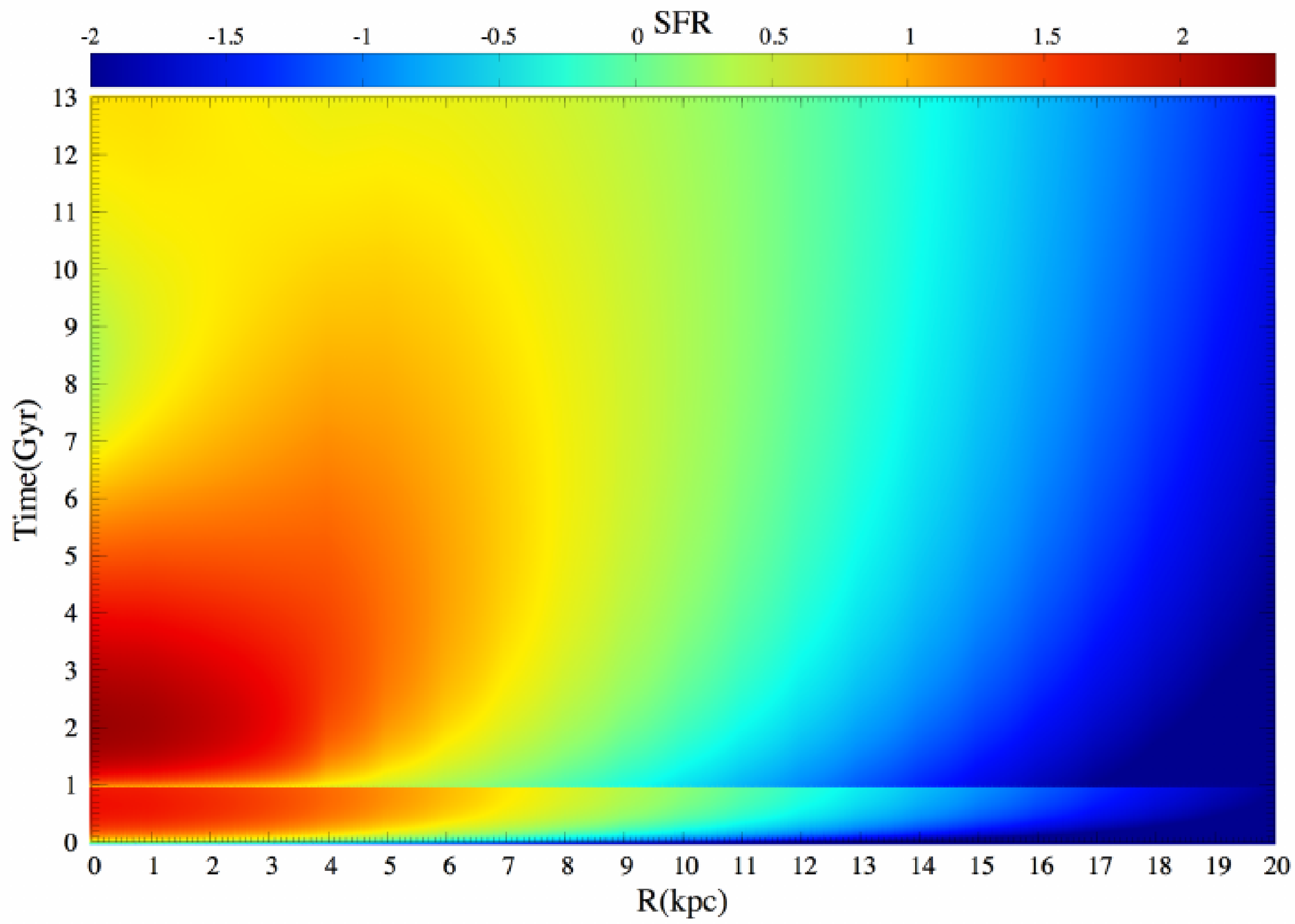}
\end{center}
\caption{SFR, along the entire disc, as a function of galactocentric
  radius and time, including an extrapolation towards the inner 3
  kpc.}
\label{f5}
\end{figure}

Before proceeding with the tagging, another key ingredient provided by
the chemical evolution model is the Star Formation Rate (SFR) across
the disc. Figures \ref{f4} and \ref{f5} show the assumed SFR for the chemical
model employed here. It is important to indicate that, as explained
before, the original SFR within the model does not provide information
for $R$ $<$ $3$ kpc, which means that we need to make a new assumption
for the inner disc.  Figures \ref{f4} and \ref{f5}  already show the
extrapolation of the SFR towards the inner 3 kpc, which was performed
as follows.

The entire stellar content of the MW disc is a compendium of a large
number of Mono Age Populations (MAPs), covering a very wide range of
ages and birth places. Because stars of different ages can coexist at
the same epoch, the current distribution of stars in the disc depends
on the distribution of individual MAPs, which in turn is dictated by
the SFR at each epoch. This means that the present time stellar
distribution depends on the SFR at all epochs. On the other hand, the
MW model employed here assumes that the stellar distribution of the
disc is well characterized by a Miyamoto-Nagai (M-N) density profile,
an assumption that works very well in reproducing key observables of
the MW \citep{AS91}.

These two previous arguments can be combined to give the main
assumption employed here, i.e., that a M-N density profile represents
the current distribution of stars in the disc, and that this
distribution is the result of the added contributions of the SFRs at
all epochs. To fulfil this constraint, and in order to extrapolate the
SFR to $R$ $<$ $3$ kpc, we took from the original data not only the
SFR at a particular epoch but the sum of the SFRs at all epochs. Next,
we fit that sum to a M-N density profile in the center. In this way
the fit of the sum give us a functional form that takes into account
all the SFRs, and now this functional expression can be used to
extrapolate, towards the center, the SFR at any time as shown in
Figure \ref{f4}.

These two physically motivated extrapolations provide us with complete
information about the chemical content (Figure \ref{f3}) and
SFR (Figure \ref{f5}) of the MW stellar disc. Now we can proceed
with the chemical tagging of the particles in our simulations, which
consists of the next steps.

The simulation goes from 0 to 13 Gyr and uses one MAP every 20 Myr,
for a total of 650 MAPs. The initial conditions for each MAP are
constructed as follows: (i) by discretizing a M-N disc profile, (ii)
the particle disc is placed inside the axisymmetric potential of the
Galactic model and, to avoid spurious transients, we evolve the system
for 0.5 Gyr while growing the bar and spirals adiabatically; after
which the system is relaxed with the background potential. (iii) Next,
to add a MAP to the simulation we take a number of particles
proportional to the SFR at every radius and at the epoch represented
by each MAP, Figure \ref{f5}. (iv) Finally, once we have the initial
positions, velocities, and the correct number of stars for every MAP,
each particle is assigned with the metallicity corresponding at its
initial radial position (birth radius) and at that specific epoch
(Figure \ref{f3}).

At the end, our simulations consist of $1.03\times10^8$ stars, with ages
between 0 and 13 Gyr, a time in which they evolved dynamically within
the galactic potential in the presence of the spiral arms and the bar.

\section{Dependence of the MDF on the Morphology of the Spiral Arms}
\label{shape}

In Section \ref{Proxy} we have shown how the shape of the initial
radii distributions changes with variations of the spiral pattern's
features, which lead us to anticipate that the MDFs will also depend
on those parameters. By tagging the stars in the simulation with the
correct metallicity, according to its birth place and epoch, now we
can show that the shape of the MDF contains information about the
spiral arms.

\begin{figure}
\begin{center}
\includegraphics[width=9cm]{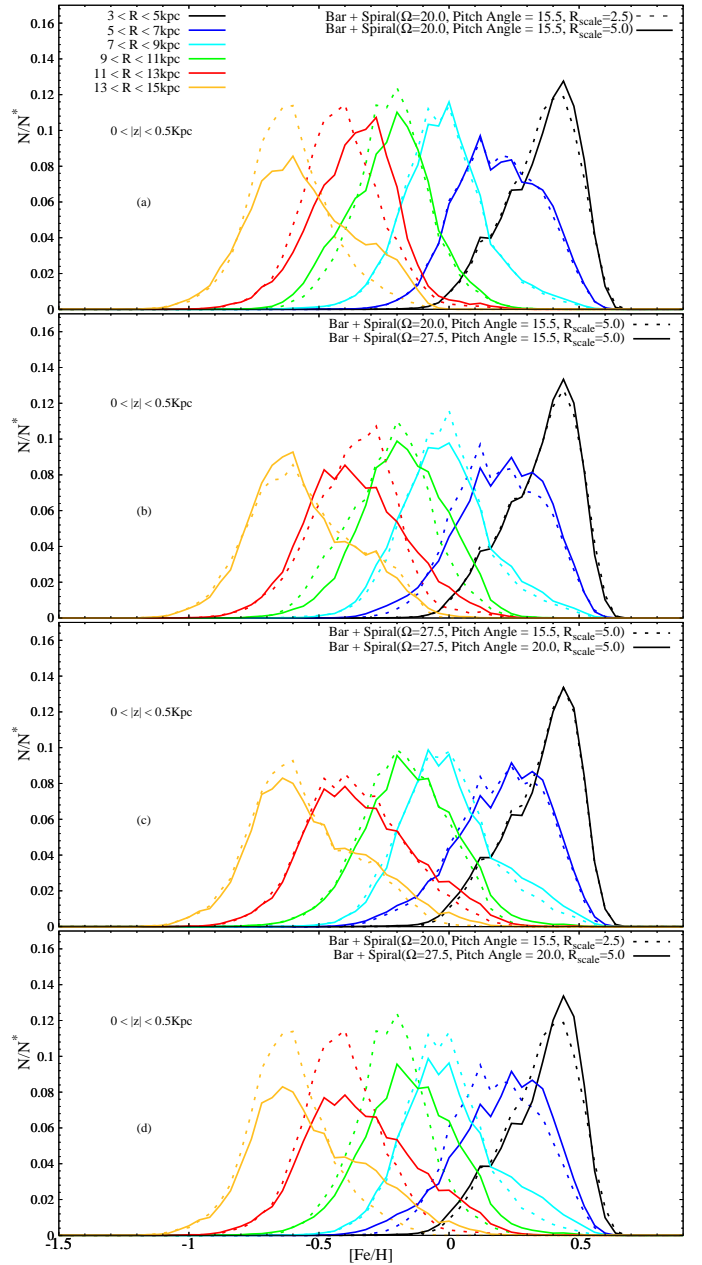}
\end{center}
\caption{MDF as a function of the galactocentric radius for different
  characteristics of the spiral arms for a population of stars between
  4 and 4.2 Gyr. The panels show how the shape of the MDF depends on
  (a) the scale radius of the spirals, (b) the pattern speed, (c) the
  pitch angle, and (d) the combination of those three parameters.}
\label{f6}
\end{figure}

From each one of our simulations we take a population of stars with
ages between 4 and 4.2 Gyr, and compute the MDFs at different radial
bins.  In Figure \ref{f6} we compare the MDF of several
galactic models that differ in the structural parameters of the spiral
arms.  Top panel shows how the shape of the MDF depends on the scale
radius of the spirals, i.e., the mass concentration of the pattern. A
less concentrated pattern means more arm's mass at outer radii, which
in turn means that stars at those locations will experience a larger
gravitational attraction during their encounters with the
spirals. Because in its orbital motion the acceleration towards the
arms will be larger, less concentrated arms will induce a greater
exchange of angular momentum and hence, a more significative radial
migration.  In the top panel of Figure \ref{f6} we can see
that the MDF is almost unaffected in the three inner radial bins. For
the rest of the bins the distributions are wider and their peaks are
lower for the less concentrated spirals, which means that the radial
mixing is larger. Notice that the radial mixing due to less
concentrated arms can be so efficient that, for example, the MDF
corresponding to the red bin ($11<R<13$ kpc) reverse its
skweness compared to more concentrated spiral arms. Also, for the
yellow bin ($13<R< 15$ kpc), there is more migration of stars
that come from the inner disc. This is reflected on a MDF with an
inner tail of higher amplitude, which increases the asymmetry of the
curve.

The most important dynamical parameter inducing migration in the outer
disc is the spiral arm pattern speed. This happens because the value
of the pattern speed determines the position of the corotation radius,
and radial migration occurs when the stars exchange angular momentum
with the spiral arms around corotation. The second panel of Figure
\ref{f6} shows the comparison of two models that differ in
the spiral pattern speed. When changing $\Omega$ from 20 to 27.5 km
s$^{-1}$ kpc$^{-1}$ the corotation radius moves from 10.9 to 8
kpc. The most affected MDF is the red one ($11<R<13$ kpc), its
skewness is reversed again and its inner tail raises, which means that
this radial bin has more contribution of stars that come from the
inner disc. The other MDF that changes significantly is the green one
($9<R<11$ kpc), being shifted towards higher metallicities.

The other important parameter of the spiral arms, that determines the
amount of angular momentum that the pattern can exchange, is the pitch
angle. An increase of the pitch angle increases the gravitational
attraction of the spiral perturbation on the stars along their orbital
direction. By accelerating stars along their rotational motion, they
gain or lose more angular momentum, moving to larger or smaller radii.
The third panel of Figure \ref{f6} shows the comparison of
two models that differ in the pitch angle of the spiral arms. By
changing slightly the pitch angle, from $15.5\deg$ to $20\deg$, we can
see that the most affected MDFs are those of the two outer radial bins
($11<R<13$ kpc and $13<R<15$ kpc). The inner tail for both
distributions raises enough to see that these radial bins are getting
more contribution of metal rich stars that were born at inner radii,
increasing the skewness of the curves.

Finally, the bottom panel of Figure \ref{f6} shows the differences in
the MDFs when we change all those three important structural
parameters of the spiral arms: scale radius, pattern speed, and pitch
angle.  The differences between both models are significant, showing
that the shape of the MDF contains information about the spiral
pattern as well as of the dynamical evolution of the stellar disc.

\section{Constraining the metallicity gradient in the inner and outer MW disc}
\label{constrain}

As mentioned in Section \ref{tagging}, chemical evolution models for
the MW disc do not include, in general, information within the inner 3
kpc. This is due to the difficulty to separate the different Galactic
components that overlap in the center of the Galaxy, but also due to
the lack of observational data to constrain a chemical model in the
central region.

A common assumption to overcome this problem is to take the value of
the chemical abundance at $R=3$ kpc, and assign it for all the inner
region ($R<3$) kpc, i.e., the metallicity gradient will be zero in the
inner 3 kpc.  Although constant chemical content in the center of the
Galaxy is a first guess, it probably is not realistic. From a
dynamical point of view, it is known that the stars can displace
radially by several kpc, driving a chemical ``contamination'' at all
galactocentric distances (i.e. different abundances from those of the
birth radius). 

From our simulations we notice that the most inner radial bin
considered ($3<R<5$ kpc) receives a significant number of stars born
at $R<3$ kpc (see for example Fig. \ref{f1}). If all the stars
born at $R<3$ kpc share the same metallicity (zero metallicity
gradient assumed), they would contaminate the stellar disc with the
same metallicity when they displace to outer radii. This would be
reflected in our two inner MDFs ($3<R<5$ kpc and $5<R<$ 7 kpc) by
producing a huge spike at a specific [Fe/H] value. This would produce
unrealistic distribution curves that show us that assuming zero
metallicity gradient in the inner disc is unsuitable. In other words,
the assumed chemical content in the inner 3 kpc will have an important
impact on the observables at $R>3$ kpc. Here we propose a fit to the
MDFs of the MW as a method to constrain the chemical evolution models
toward the center of the disc.

Noticing that close to the plane, the two inner MDFs reported by
\citet[][Fig. 5]{Hayden2015} overlap, we tried to obtain similar MDFs
from our simulations by changing the gradient of the chemistry curves,
at R $<$ 3 kpc, in order to avoid the unrealistic spike in the inner
MDFs, tuning the gradient to obtain a similar degree of overlapping,
as is observed. Our best fit to the observed MDFs provide us a
chemical curve along the entire disc that includes the prediction at
inner radii in Figure \ref{f3}.

Moreover, since the observed MDFs peak at high metallicities in the
inner Galaxy, then peak approximately at a solar value in the solar
neighbourhood, and peak at lower metallicities in the outer disc, the
observed MDFs also show the radial metallicity gradient across the
disc, which puts additional constraints on our simulation. When using
the chemical gradient as presented in Figure \ref{f3},  we find 
that the stars lie between $-0.65 <$ [Fe/H] $< 0.65$, while the
vast majority of stars in the sample presented by \citet{Hayden2015}
have values of [Fe/H] between $-0.5$ and 0.5. This suggests that the
actual chemical gradient of the MW is slightly flatter than the
$-0.102$ dex kpc$^{-1}$ presented in Figure \ref{f3}. By flattening
the chemical gradient by 22\% to $-0.080$ dex kpc$^{-1}$, we reproduce
the range observed by Hayden et al. This means that, by confining the
peaks of the simulated MDFs to the reported values, we improve the
chemical evolution model via the restriction of the radial gradient
across the disc.

\begin{figure}
\begin{center}
\includegraphics[width=9cm]{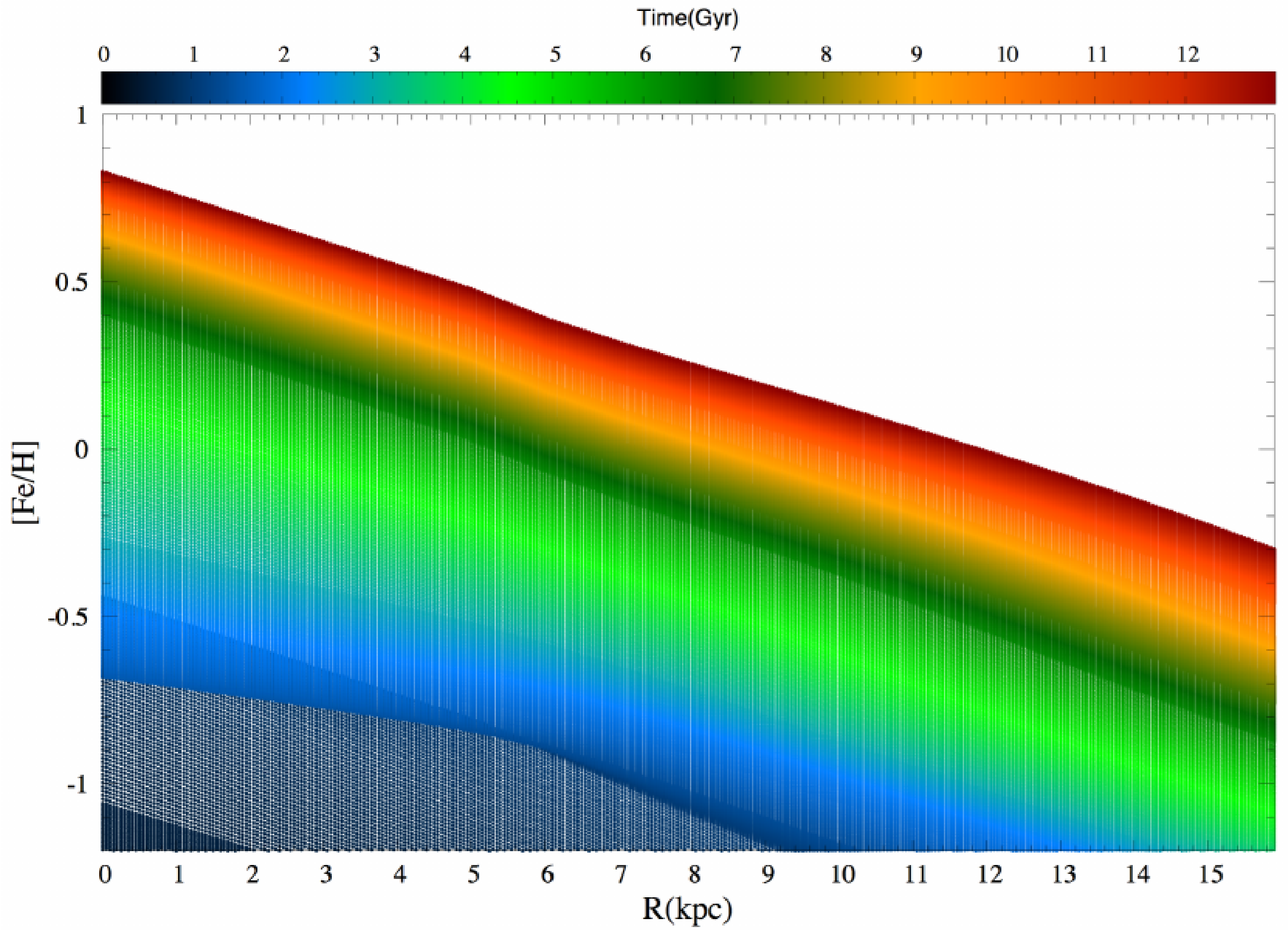}
\end{center}
\caption{[Fe/H] curves for the chemical evolution model of the MW,
  including the extrapolation towards the inner 3kpc and the modified
  gradients to better reproduce the resulting MDFs. The color code
  traces the history of the Galactic disc, from its formation to the
  present time (at 13 Gyr).}
\label{f7}
\end{figure}

Figure \ref{f7} shows the final chemical model obtained by (i)
extrapolating the chemical content towards the inner disc (in order to
overlap the two inner MDFs, without the unrealistic spike), and (ii)
confining of the MDF peaks between $-0.5<$ [Fe/H] $<0.5$.

\section{Stellar dynamics imprinted in the Metallicity Distribution}
\label{stellardynamics}

In this work we propose an analysis of the MDFs as a method to obtain
information on the structure and dynamics that shape the spiral arms
and bar of galaxies. We apply this to the Milky Way, where the MDFs
have been observed in great detail.

As we have mentioned, the interactions of the stars with the
non-axisymmetric structures in the disc can modify the stellar orbits
either by heating or by radial migration. Heating occurs when the
orbital eccentricity increases, and can be quantified directly from
the kinematics of the stars; in this case, the velocity dispersion can
provide a good idea of the dynamical evolution of a given star field;
e.g. it is known that the age and velocity dispersion of stars are
correlated \citep[see for
  example][]{Holmberg2009,Roskar2013,Gerssen2012,2007MNRAS.380.1348S,1974A&A....32..321M,1985ApJ...294..674C}. On
the other hand, when migration takes place the mean galactocentric
distance of the star changes significantly, but its orbital
eccentricity does not; this means that, unlike with heating, migration
cannot be revealed through stellar kinematics; instead, recent
observations and numerical simulations have shown that this dynamical
process can be revealed through the MDFs.

Since the shapes of the MDFs measure displacements of stars away form
their birth radius, they capture both processes, heating as well as
migration. We argue that the shape of the MDF not only tells us
whether the stellar disc experienced migration, but also contains
specific information about the structure that triggered such
migration. We have come up with this conclusion by (i) finding that
the spiral arms are the ones that determine the sign in the skweness
of the MDFs \citep{MartinezMedina2016}, and (ii) showing how the
shapes of the MDFs are sensitive to the structural parameters of the
spiral pattern (see Section \ref{shape}).

\begin{figure}
\begin{center}
\includegraphics[width=9cm]{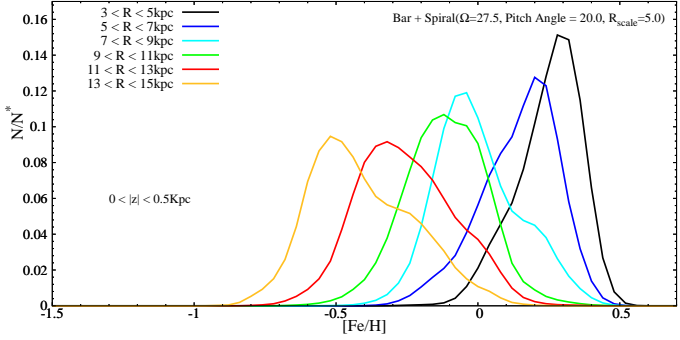}
\end{center}
\caption{MDFs as a function of galactocentric radius for a Galactic
  model with the parameters of Table \ref{tab:parameters} and iron
  abundance gradients from Figure \ref{f7}, for stars with
  ages between 4.6 and 5.6 Gyr.}
\label{f8}
\end{figure}

In Figure \ref{f8} we present the results of a simulation
that includes the central bar and the spiral pattern (see Table
\ref{tab:parameters} for the specific parameters employed). In this
figure, we have reproduced closely characteristics observed in the
MDFs of the MW, as reported by \citet{Hayden2015}, such as, (i) the
negative skweness in the inner disc (with tails that extend to low
metallicities), which is reversed at the outer disc (with tails that
extend to high metallicities); (ii) there is an evident radial
metallicity gradient across the disc in spite of the significant
radial mixing; and (iii) the peaks of the MDFs are confined within the
interval $-0.5<$ [Fe/H] $<0.5$, which provides us the value of the
radial gradient.

\begin{table}
\centering
\caption{Parameters of the non-axisymmetric Galactic components.}
\label{tab:parameters}
\begin{tabular}{lc}
 \hline
 \hline

Parameter & Value \\
 \hline

\multicolumn{2}{c}{\it Triaxial ellipsoidal Bar}\\
 \hline
Major Semi-Axis                     & 3.5 kpc         \\
Scale Lengths                       & 1.7, 0.64, 0.44 kpc \\
Axial Ratios                        & 0.64/1.7, 0.44/1.7  \\
Mass                                & 1.4 $\times$ 10$^{10}$ M$_{\odot}$  \\
Pattern Speed ($\Omega_B$)          & 45$\kmskpc$ \\
 \hline
\multicolumn{2}{c}{\it Spiral Arms}\\
 \hline
Number of spiral arms                   & 2 \\
Pitch Angle ($i$)                   & 20$^{\circ}$ \\
Radial Scale Length ($H_{\star}$)          & 5 kpc         \\
$M_{\rm arms}/M_{\rm disc}$         &  0.05  \\
Mass                                &  4.28 $\times$ 10$^9$ M$_{\odot}$  \\
Pattern Speed ($\Omega_S$)          & 27.5 $\kmskpc$  \\
 \hline
\end{tabular}
\end{table}

\begin{figure}
\begin{center}
\includegraphics[width=9cm]{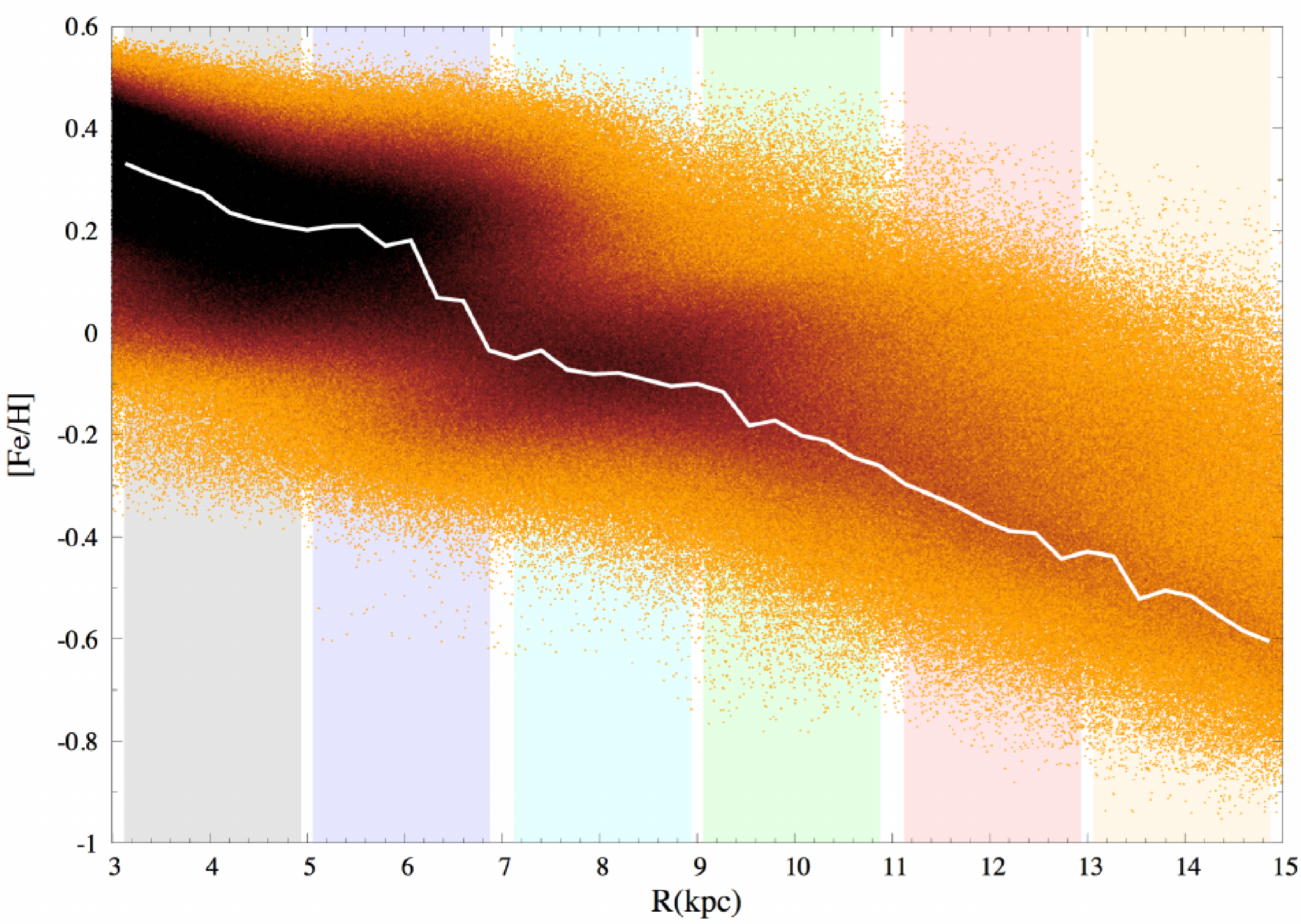}
\end{center}
\caption{Distribution of stars in the plane [Fe/H] vs. Radius
  corresponding to the simulation of Figure \ref{f8}, and stellar ages between 4.6 and 5.6 Gyr.
  The colour code indicates density of stars, while the solid white
  line is the mode of the [Fe/H] values at every radius. The
  background coloured bands are set as reference to compare with the
  same color MDFs in Figure \ref{f8}.}
\label{f9}
\end{figure}

Heretofore, the analysis of the MDF gave us important clues on the
chemical and dynamical evolution of the stellar disc. A more detailed
insight of what is happening in the disc can also be obtained by
studying the plane [Fe/H] $vs.$ $R$. Figure \ref{f9} shows the
metallicity distribution across the disc for a population of stars
with ages between $4.6$ and $5.6$ Gyr, in a MW model with the
parameters for the bar and spiral arms given on Table
\ref{tab:parameters}. Every point in the plot represents a star at its
current galactocentric radius and its metallicity, the color coding
indicates the density of stars. In Figure \ref{f9} we can see in more
detail the extended tails that appear in the MDFs from previous
figures; for instance, at the inner radial bins the peak in density of
stars is not centered in the middle of the orange band, but a little
shifted to the top; on the other hand, in the outer radial bins, the
peak in density is located at low metallicities, near the bottom of
the orange band, with an important number of stars at high
metallicities. Also, while the MDF tells us the radial bin to which
the star belongs, the [Fe/H] $vs.$ $R$ plane provides its exact
location within the bin. For example, stars with high metallicity that
were born at the inner disc but now are located in the outer radial
bin ($13<R<15$ kpc) did not migrate to only reach the inner boundary
of that bin, but as Figure \ref{f9} shows, they migrated also to the
outer boundary of the bin.

The most prominent features in Figure \ref{f9} are the two
overdensities at $R=4.7$ kpc, [Fe/H]=0.25 and at $R=8$ kpc,
[Fe/H]=-0.1. The break between these two overdensities marks the
transition between the influence of the bar to the influence of the
spiral pattern, i.e. the inner overdensity is composed by stars that
have been strongly affected by the corotation of the bar, while the
outer one is composed by stars that have been strongly affected by the
corotation of the spiral pattern. This behavior is studied in depth in
\citet{MartinezMedina2016}.

\begin{figure}
\begin{center}
\includegraphics[width=9cm]{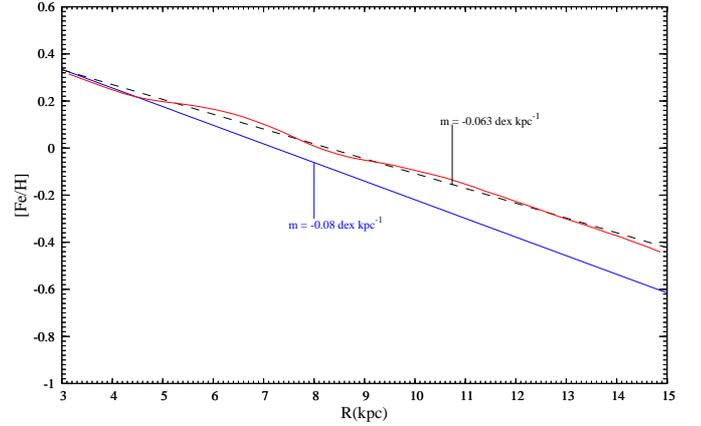}
\end{center}
\caption{The mean of the [Fe/H] values (red line) for the data plotted
  in Figure \ref{f9}. The blue line indicates a linear fit to
  the iron abundance of the stellar population at its birth epoch.
  Black dashed line is a linear fit to the mean iron abundance at the
  present time.}
\label{f10}
\end{figure}

Notice that Figure \ref{f9} shows the density of stars in the [Fe/H]
$vs.$ $R$ plane, and hence the mode of the data (white curve)
indicates directly the density peaks. However, to quantify the
metallicity gradients for this stellar population, Figure \ref{f10}
shows the mean of [Fe/H] as a function of $R$.  The blue line
corresponds to the initial metallicity gradient, which is a linear fit
to the iron abundance of the population at its birth epoch. The red
curve is the mean of the data plotted in Figure \ref{f9}, while the
black dashed line is a linear fit to the red curve, and hence, the
metallicity gradient of the present stellar population.  Its clear
that the metallicity gradient evolved with time, but still the two
slopes are not very different. It is worth noticing that at the
present time the stellar population exhibits a clear metallicity
gradient in spite of the important radial migration and heating in the
disc. This means that the presence of important radial migration does
not imply necessarily a substantial flattening of the metallicity
gradient (see also \citet{2017MNRAS.tmp..124S} for a similar
conclusion).

\subsection{Metallicity gradients away from the Galactic plane}

Because stars with small vertical oscillations remain close to the
plane, having more interaction with the non-axisymmetric structures on
the disc, they are the most efficiently affected by radial migration.
However, stars with large vertical excursions or large vertical
velocities still can experience radial displacements, that could contribute 
to a changing metallicty gradient when the altitude away 
from the plane increases, as is the case for the solar neighborhood 
\citep{2004A&A...418..989N, 2013A&A...559A..59B, 2014A&A...568A..71B}.

\begin{figure}
\begin{center}
\includegraphics[width=9cm]{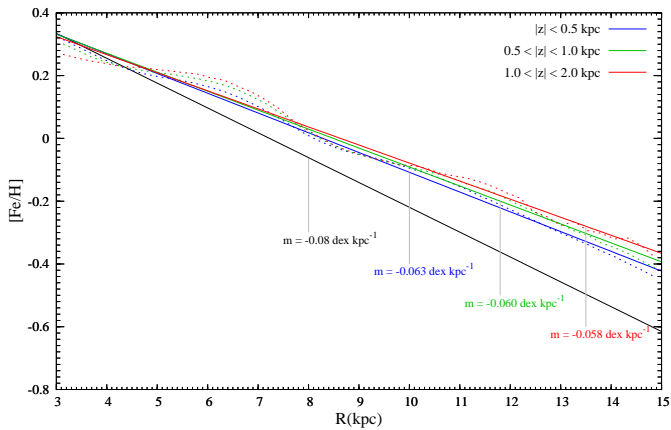}
\end{center}
\caption{Radial metallicity gradients, for a stellar population with
  average age of 5Gyr, at different vertical distances from the
  plane. The black line indicates the initial gradient for the entire
  population regardless of $z$.}
\label{f11}
\end{figure}

In this section we go beyond the thin disc to compare the metallicity
gradient obtained at different ranges of $z$.  As mentioned above,
from our simulation we took a population of stars with ages between
$4.6$ and $5.6$ Gyr.  Figure \ref{f11} shows the mean of
the [Fe/H] distribution as a function of $R$ for three ranges of $z$
(dashed lines). The solid lines indicate a linear fit for each one of
these mean curves, while the black line is the initial [Fe/H] gradient
of the entire population.  The first thing to notice is that the final
metallicity gradient becomes less steep with increasing
height. However, although the tendency is clear, the differences
between the three gradients is rather small.
  
We interpret the trend shown in Figure \ref{f11} not as a
direct consequence of radial migration, but as a result of the biased
nature of the mechanism. In other words, because the stellar density
of the disc decreases with radius, more migrators will arrive, at a
given radial positon, from the inner disc than from the outer disc;
this means that at that radial position the excess of metal-rich stars
will contribute positively to the metallicity gradient, making it less
steep. The effect becomes evident with increasing radius and even more
with increasing height, where the number of stars is
small. \citet{2017MNRAS.464..702K} obtain a similar result, observing
that this bias leads to a positive vertical metallicity gradient.

\begin{figure}
\begin{center}
\includegraphics[width=9cm]{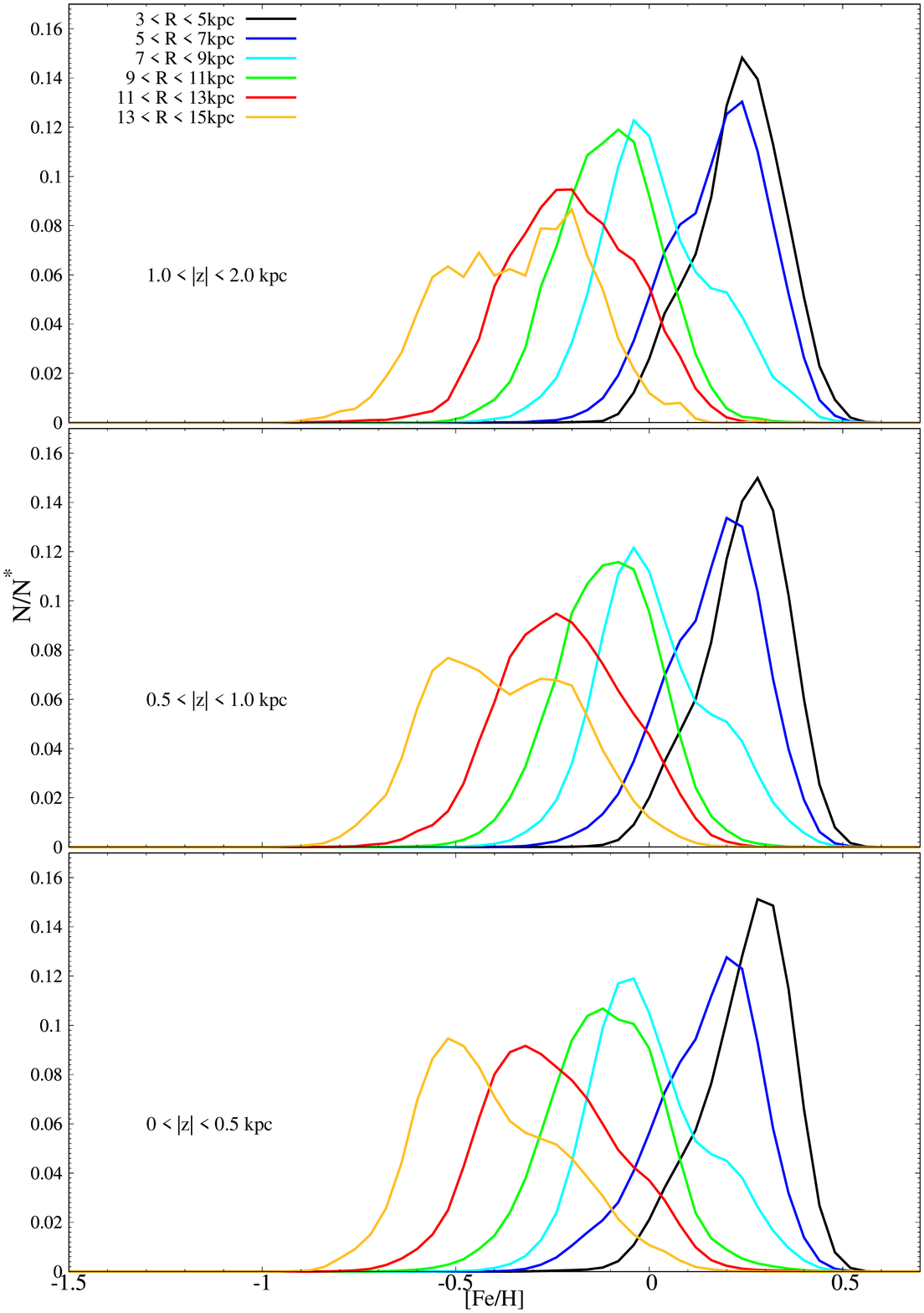}
\end{center}
\caption{MDFs as a function of galactocentric radius, at different
  heights away from the plane, for stars with ages between 4.6 and 5.6
  Gyr. The computations were performed using a Galactic model with the
  parameters of Table \ref{tab:parameters} and iron abundance
  gradients from Figure \ref{f7}.}
\label{f12}
\end{figure}

The bias in the radial migration mechanism, producing more outward
than inward migrators, is also reflected in the MDFs.  Figure
\ref{f12} shows the MDFs across the disc, going in $|z|$ from 0
to 2 kpc. Notice that when $|z|$ increases, the position of the peaks
and the shape of the curves remain nearly unchanged, except for the
two outer radial bins (red and yellow distributions). The peak of the
outmost MDF shifts to higher metallicities as the height increases,
just reflecting the fact that at those radial positions and altitudes
there are still stars arriving from the inner disc, but the density
has dropped to a point that there are almost no stars coming from the
outer disc.

At the end, although this bias affects mainly the outer regions of the
disc and becomes more evident at high altitudes, Figure
\ref{f11} shows that a metallicity gradient is clear and
differs just slightly when going from the thin to the thick disc.

\section{The radial reach of migration and heating}
\label{radialreach}
In the previous sections we have seen how the presence of spiral arms
or a bar in a disc galaxy displace stars radially, with an important
fraction of them being moved away from their birth radius. As already
mentioned, the displacements occur because the stars gain or loose
angular momentum when interacting with the pattern (spirals or
bar). For example, in the case of spiral arms, a star can be
orbitating either in front of the pattern or behind it.  If the star
is ahead of the arm, the overdensity of the pattern will pull the star
backwards, diminishing its angular momentum, which will cause the star
to move to inner radii. Similarly, if the star is located behind the
pattern, it will be pulled forward, which increases its angular
momentum, moving the star to larger radii. These two scenarios mean
that, at any given radius, the spirals or the bar can induce radial
displacements in both directions.

\begin{figure}
\begin{center}
\includegraphics[width=9cm]{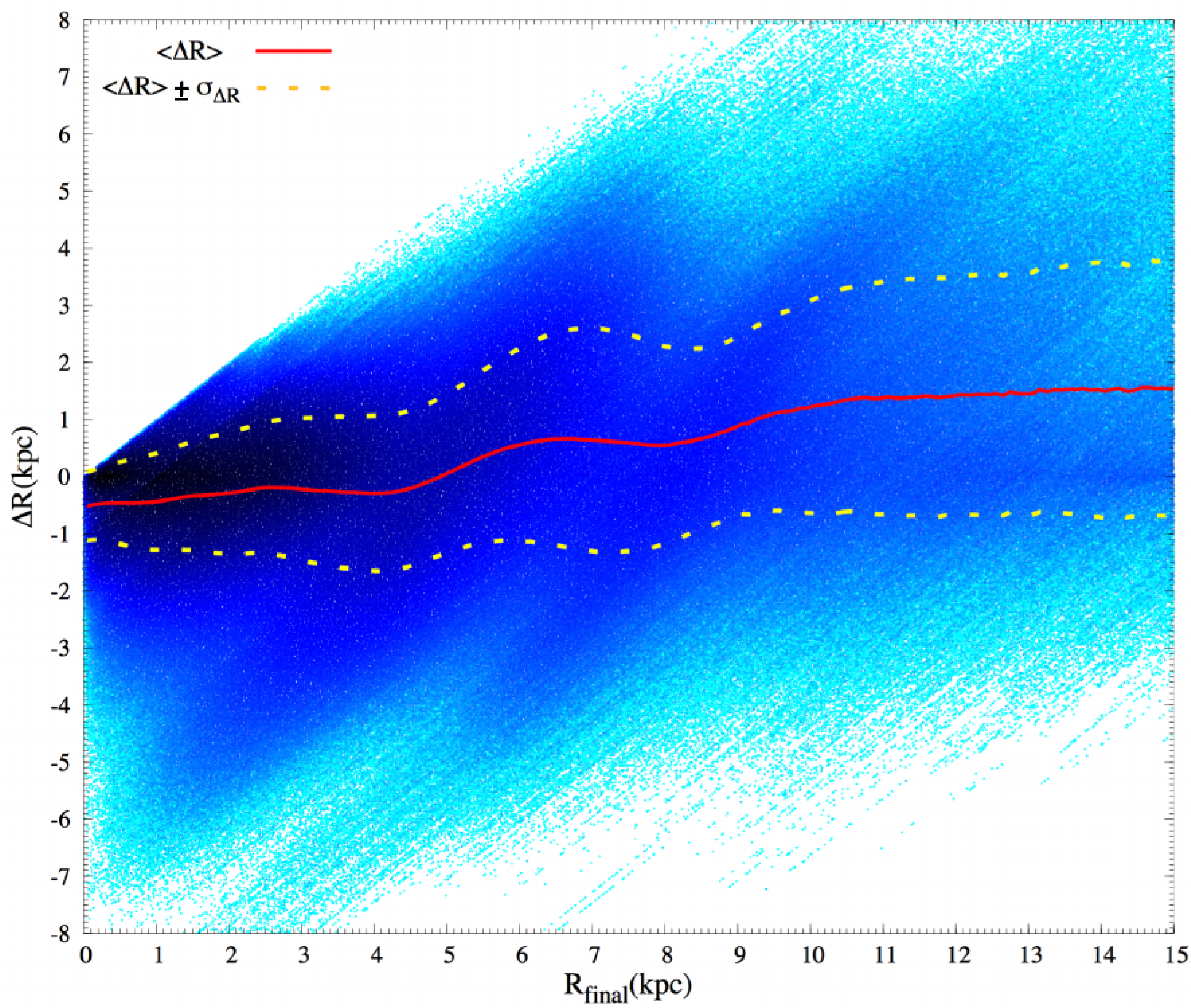}
\end{center}
\caption{Difference between the current position of every star and its
  birth radius, $\Delta R = R_{final} - R_{birth}$, as a function of
  its current galactocentric position $R_{final}$, for a population
  with ages between $4.6$ and $5.6$ Gyr. The color coding indicates
  the density of stars in logarithmic scale. The red solid line shows
  the mean of the $\Delta R$ values as a function of $R_{final}$. The
  dashed lines indicate the dispersion of the data,
  $\pm\sigma_{\Delta R}$ around the mean, as a function of $R_{final}$.  }
\label{f13}
\end{figure}

In this section we quantify the extension of the radial displacements
in the orbits of stars.  Figure \ref{f13} shows the difference
between the current position of every star and its birth radius,
$\Delta R = R_{final} - R_{birth}$, as a function of its current
galactocentric position $R_{final}$, for a population with ages
between $4.6$ and $5.6$ Gyr. Notice that during their lifetime, after
evolving in the Galactic potential, an important fraction of stars end
up several kpcs away from their birth radius, with radial excursions
either to larger or inner radii.  To identify any trend in the
displacements, the red solid line in Figure \ref{f13} indicates the
mean of the $\Delta R$ values as a function of $R_{final}$. Notice
that $<\Delta R>$ is negative for the first 5 kpc, indicating a
tendency to move from the outer to the inner disc. Meanwhile for
$R_{final}>5$ kpc the trend is that, in average, the stars were born
at inner radii, and were displaced outwards to their current position.

In the particular case of the solar circle ($R_0 = 8.5$ kpc) the
average displacement is $<\Delta R>_0 \approx 0.68$ kpc. However, this
value is just the average, and the fraction of stars that were born at
radii considerably smaller and end up at the solar radius can be
important.  The dashed lines in Figure \ref{f13} indicate the
dispersion of the data, $\pm\sigma_{\Delta R}$ around the mean, as a
function of $R_{final}$. At the solar circle $\sigma_{\Delta R}
\approx 1.57$ kpc, which added to $<\Delta R>_0$ means that stars
within 1$\sigma_{\Delta R}$ reached the solar circle from
galactocentric positions $6.25 < R_{birth} < 9.39$ kpc. Table
\ref{tab:DR} shows the same analysis for different galactocentric
distances across the disc. Within a 1$\sigma_{\Delta R}$ dispersion,
stars with $R_{final} > 6$ kpc can be displaced for more than 2
kpc. Figure \ref{f13} also shows that the radial displacements for
$R_{final}>6$ kpc can be considerably large.

\begin{table}
\centering
\caption{Radial displacements accross the disc.}
\label{tab:DR}
\begin{tabular}{ccc}
 \hline
 \hline
$R_{final}$(kpc) & $<\Delta R>$(kpc) & $\sigma_{\Delta R}$(kpc) \\
 \hline
1 & -0.43 & 0.85\\
2 & -0.28 & 1.07\\
3 & -0.22 & 1.25\\
4 & -0.29 & 1.36\\
5 & 0.08 & 1.4\\
6 & 0.57 & 1.7\\
7 & 0.64 & 1.9\\
8 & 0.55 & 1.7\\
8.5 & 0.68 & 1.57\\
9 & 0.9 & 1.57\\
10 & 1.24 & 1.88\\
11 & 1.4 & 2\\
12& 1.43 & 2.1\\
 \hline
\end{tabular}
\end{table}

\subsection{The Birth Place of the Sun}
We have seen that stars can undergo radial displacements of several
kiloparsecs, meaning that for an important fraction of stars their
birth radius is very different from their current galactocentric
distance. Moreover, during its evolution in the galactic potential,
the orbital elements of a star can change rapidly, erasing all
kinematic hints about their birth radius. However, as we already
mentioned, an indication of whether the stars have migrated or not,
and how far are they from their birth radius, is their chemical
content.

For the particular case of the Sun, it has been found that it is more
metal rich than the majority of solar age stars in the solar
neighbourhood. The high metallicity of the Sun can be explained
assuming that it migrated outwards, from its birth, place to its
current galactocentric distance
\citep[e.g.][]{1996A&A...314..438W,Holmberg2009,2012A&A...539A.143N}.
  Different results, based only on orbital dynamics, show a possible
  migration inwards, from its birth place, to its current position
  \citep{2015MNRAS.446..823M}. In this section we discuss plausible
  sources of discrepancy with chemical determinations and with the
  results in this work.

Taking advantage of the large number of particles in our simulations,
we test the plausibility of a migration scenario for the Sun. From the
simulation we select the stars with orbital properties similar to the
current ones of the Sun, i.e, position, velocity with respect to the
local standard of rest (LSR), and age. Then, for those stars that
fulfill the selection criteria, we trace back their birth radii.

For the present-day galactocentric distance of the Sun, its altitude
above the galactic plane, and its velocities, we take the values
listed in Table \ref{tab:SunParameters}. The velocities
$(U,V,W)_{\odot}$ \citep{2010MNRAS.403.1829S} are taken with respect
to the LSR, that in our galactic model has the value $V_{LSR}$ = 220
km s$^{-1}$. The standard deviation in each orbital parameter
corresponds to the uncertainties in each reported value.

\begin{table}
\centering
\caption{Sun's orbital parameters.}
\label{tab:SunParameters}
\begin{tabular}{ccccc}
 \hline
 \hline
Coordinate &  & & value &  \\
 \hline
$R_{\odot}$ & & 8.5&$\pm$&0.5 kpc\\
$z_{\odot}$ & & 0.02&$\pm$&0.005 kpc\\
U & & 11.1&$\pm$&1.23 km s$^{-1}$\\
V & & 12.24&$\pm$&2.1 km s$^{-1}$\\
W & & 7.25&$\pm$&0.62 km s$^{-1}$\\
 \hline
\end{tabular}
\end{table}

\begin{figure}
\begin{center}
\includegraphics[width=9cm]{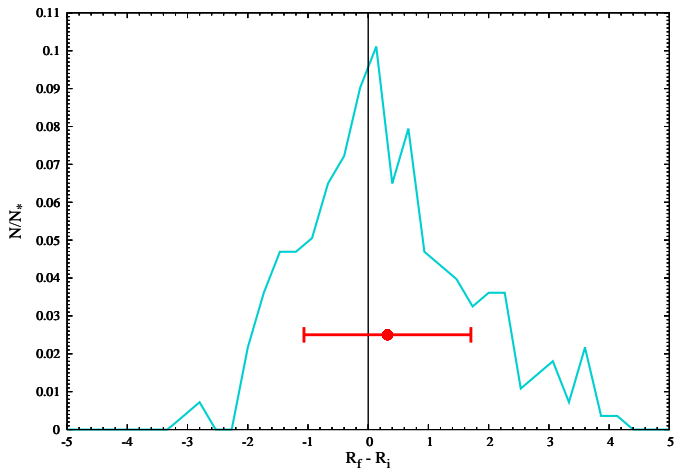}
\end{center}
\caption{Distribution of radial displacements of the stars that
  currently share the age and orbital parameters of the Sun. The
  vertical line divides inwards from outwards migrators, while the red
  dot represents the 0.32 kpc average, with a 1.38 kpc dispersion of
  the sample.}
\label{f14}
\end{figure}

\begin{figure}
\begin{center}
\includegraphics[width=9cm]{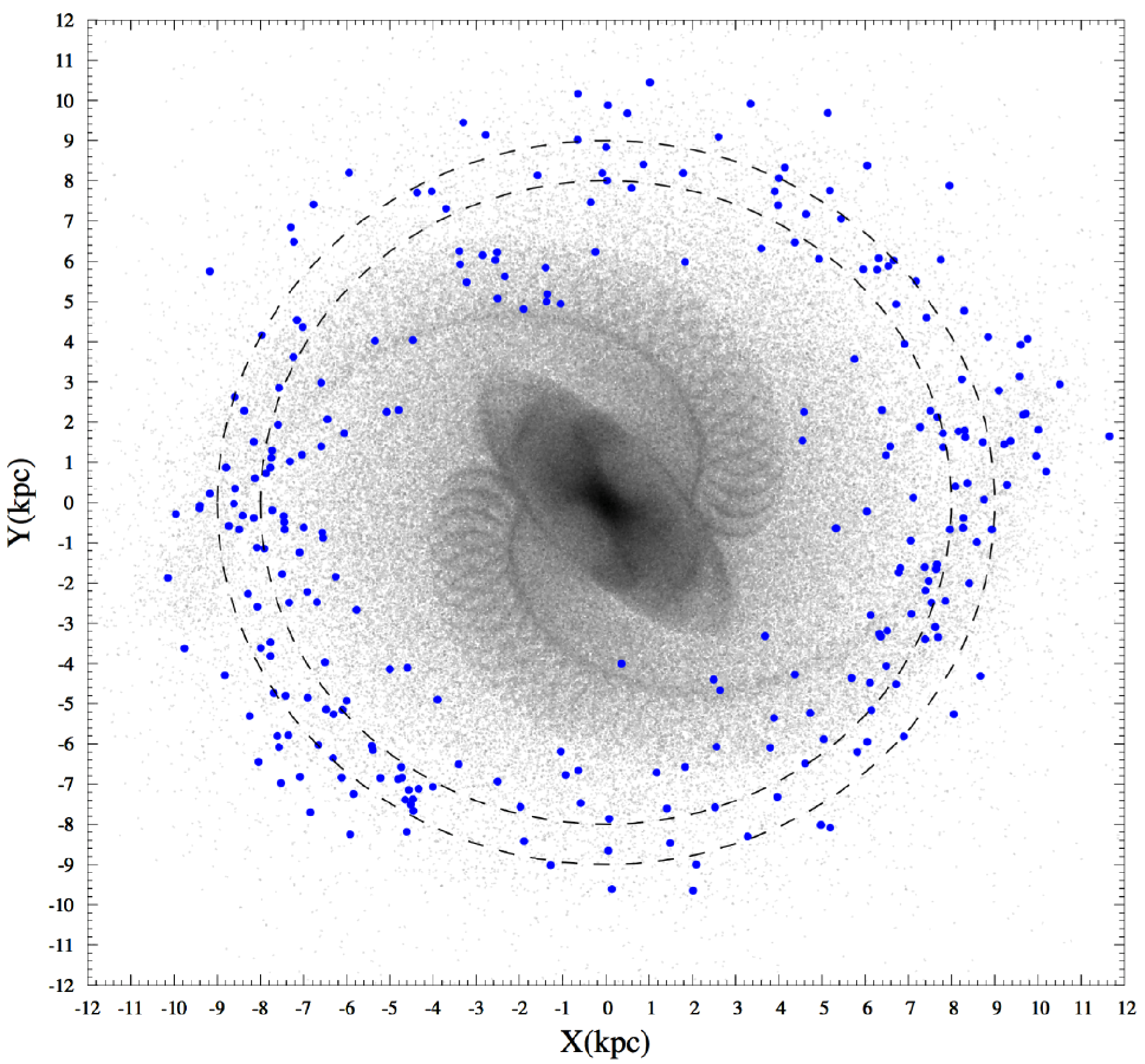}
\end{center}
\caption{Position in the galactic plane for stars that fullfilled our
  selection criteria. Blue dots indicate their birth place, while the
  gray color gradient shows the configuration of the bar and spirals
  at that moment. The ring indicates the current solar radius of $8.5
  \pm 0.5$ kpc}
\label{f15}
\end{figure}

As the mean age reported for the Sun is 4.6 Gyr, we selected those
stars with ages between 4.4 and 4.8 Gyr. When looking in our sample
for stars with orbital parameters within the restrictions presented in
Table \ref{tab:SunParameters}, we found none of our $10^8$ stars met
the criteria. For this reason, from the stars with those ages and
radii, we selected the ones with altitude above the plane $z =
z_{\odot} \pm 3\sigma_z$, and velocities $(U,V,W) = (U_{\odot} \pm
3\sigma_U, V_{\odot} \pm 3\sigma_V, W_{\odot} \pm 3\sigma_W)$, where
the standard deviation $\sigma$ was taken as the uncertainty in each
reported value. With this relaxed criteria we found only 277 stars
that fulfill all our restrictions; the scarcity of these particles is
not surprising given the amount of phase space volume available. The
result of this selection procedure is shown in Figures \ref{f14} and
\ref{f15}.

Figure \ref{f14} shows the distribution of radial displacements of our
Sun-like stars. The displacements are centered at $R_f-R_i=0.32$, with
a dispersion of 1.38 kpc; some stars migrating as much as 4 kpc
outwards or 3 kpc inwards, with 56\% of our sample moving outwards and
44\% moving inwards.

Figure \ref{f15} shows the birth place on the galactic plane as well
as the configuration of the bar and spirals 4.6 Gyr ago, for the stars
that fulfill our selection criteria. Notice that the distribution of
stars is not axisymmetric, with most stars being born in the arm or
just behind it, and very few being born ahead of the spiral arms.

The previous analysis shows plausible scenarios that place the birth
radius of the Sun away from its current position, showing, with a high
probability, that the Sun was born at a galactocentric distance of
$R_{\odot} = 8.18 \pm 1.38$ kpc. However, the fact that we find
hundreds of strongly migrating stars that mimic closely the current
solar orbital parameters, shows that the study of radial migration
will not be able to accurately constraint the birth place of the Sun;
this is true for our approximation to the Galaxy, as well as it would
be to any model of the Galaxy that is more complete (complex) than
ours. Also, the likelihood of a strong migration would pose serious
difficulties for instance, to the possibility of finding any solar
siblings since the long excursion through the Galaxy, would exert a
strong shear that would have probably dissociated the Sun's birth
cluster rapidly.

Other pure dynamical based works find a tendency for the Sun to
  have originated on the outer parts of the Galaxy
  \citep{2015MNRAS.446..823M}, this is not compatible with what is
  found chemically. However, in that work, although their conclusions
  indicate that the Sun comes from the outer parts of the Galaxy,
  their figures cannot rule out the possibility that it could have
  come from the inner region. Still, some (small) discrepancy remains
  on the percentage of migrators, relative to what it is obtained in
  this work; this could be probably due to the simplified potential
  model they employ or to the fact that integrating backwards in time
  in chaotic regions (e.g. near overlapping resonances), may introduce
  systematic errors.

Considering that it is unlikely that the Sun was formed near its
current galactocentric position, it may be inaccurate to consider the
Sun as representative of the chemical abundances in the solar
neighborhood; this was also noticed by \citet{2012A&A...539A.143N}
with chemical studies of nearby B stars. \citet{1993A&A...275..101E}
and \citet{1996A&A...314..438W} also find that the metallicity of the
Sun suggests that it was formed in the inner disk of the Galaxy.

A better way to estimate the birth radius of the Sun, is to combine
the dynamical information with what can be obtained by comparing its
chemical composition with that of young objects (HII regions, B stars,
etc.)  and, using what is known of the metallicity gradient and
chemical evolution of the MW, to determine the likely origin of the
Sun. In our chemical model, the solar metalicity, 4.6 Gyr ago, was
observed at a galactocentric distance of 7.7 kpc, and once including
reasonable errors for the Sun's chemical abundance, as well as errors
in the determination of the chemical gradient and evolution of the MW,
this translates to: $7.7 \pm 1$ kpc; combining this with our dynamical
determination we find that the Sun was likely to have formed at a
galactocentric distance of $7.9 \pm 0.8$ kpc.

\section{The MDF for different Age Populations}
\label{MAPs}

Up to this point we have shown how the MDF looks for a given MAP, or
more precisely, for stars within a narrow age range. For example, the
MDFs in Figure \ref{f8} reveal a metallicity gradient across the disc,
with long extended tails producing asymmetric curves, and with peak
metallicity values constrained between $-0.5<$ [Fe/H] $<0.5$. But how
do they look like for younger or older stars? i.e., how are the MDFs
be for different stellar ages?

As described above, our simulations are an extensive compendium of
mono-age populations, with number of stars and distribution across the
disc given by the SFR at their birth epoch. Also, as we applied a
chemical tagging according to the data in Figure \ref{f7}, we
can compute the MDFs for any age.

\begin{figure}
\begin{center}
\includegraphics[width=9cm]{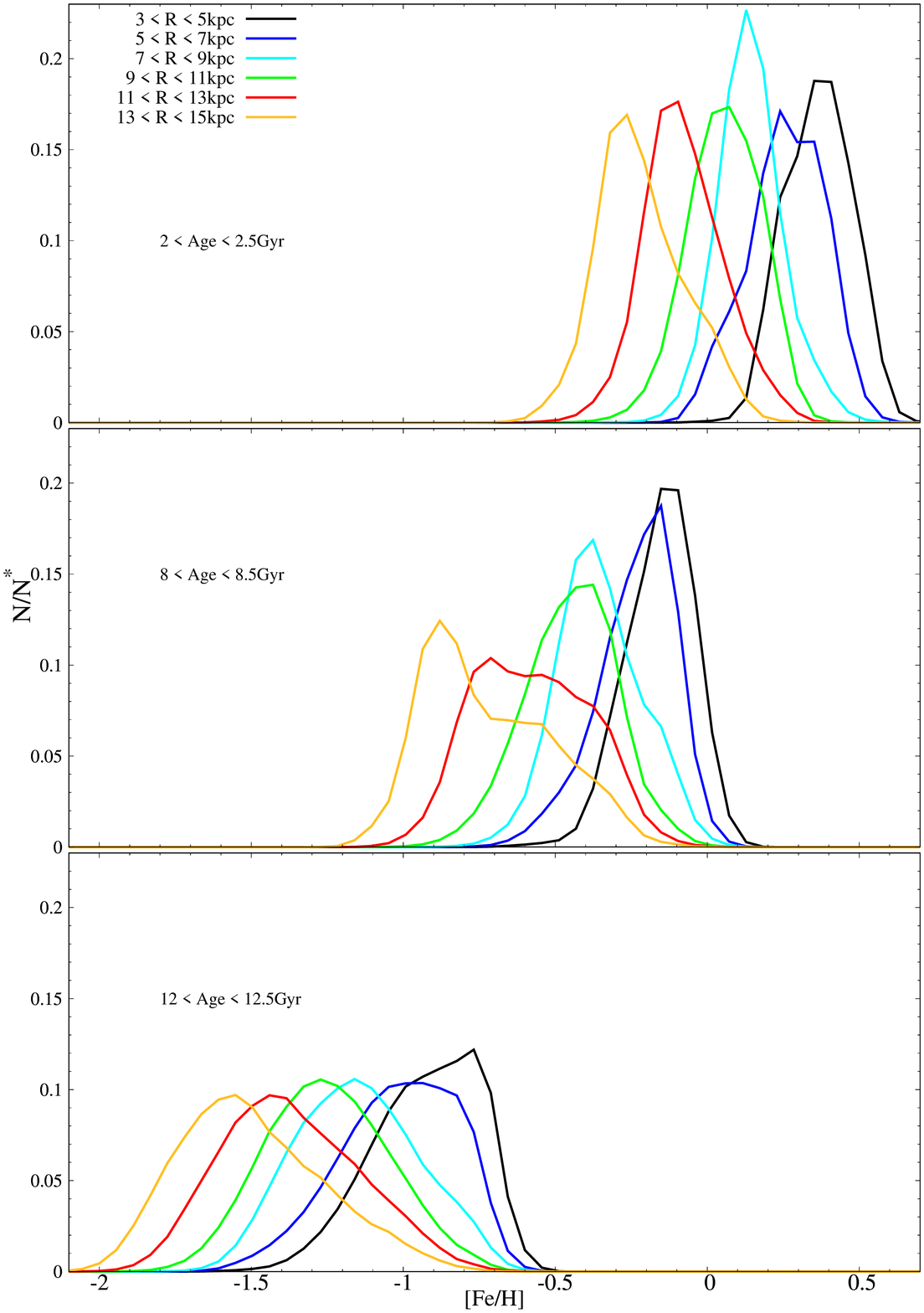}
\end{center}
\caption{MDFs for three different age intervals. \textit{Top}: for
  stars with $2<$Age$<2.5$ Gyr.  \textit{Middle}: stars with
  $8<$Age$<8.5$ Gyr. Bottom: stars with $12<$Age$<12.5$ Gyr.  }
\label{f16}
\end{figure}

Figure \ref{f16} shows the MDFs for three narrow age intervals in the
simulation, going from very young to very old stars.  In the top
panel, for the youngest stars, the MDFs are narrow with high
amplitude, meaning that, for most of the stars, the metallicity is the
one expected at that radial position. In other words, this is an
indication that the current galactocentric distance of the star is not
very different from its birth radius.

In the middle panel of Figure \ref{f16}, for stars of
intermediate age, the MDFs are clearly different from the previous
case, specially for the outer radial bins. First notice that, as a
consequence of the lower metal content of the disc at earlier epochs,
all the curves are shifted toward lower values of [Fe/H]. They are
wider, have decreased in amplitude, and the asymmetry is more
evident. These features are the signatures of an effective radial
mixing of the stellar disc, particularly one driven by radial
migration, that we know now is the responsible for the extended tails
and opposite skewness (when going from the inner to the outer disc) in
the MDFs.

The bottom panel of Figure \ref{f16} shows the MDFs for very old
stars. As in the previous case, the curves are shifted toward even
lower values of [Fe/H]; the curves have smaller amplitudes and are
wider compared to the previous cases. Note how for the inner radial
bins, the internal tail of the curve extends toward low metallicities,
while for the outer radial bins, the external tail of the curves
extends toward large metallicity values. The extension of the tails
and the pronounced asymmetry in the MDFs of very old stars is clear
evidence of significant radial mixing, which is what one would expect
from an old, dynamically more evolved population. 

We have also compared our results with \citet{2011A&A...530A.138C}
(cyan curves of our Figure \ref{f16} {\it vs.} the top panel of
Casagrande's Figure 16). While we also see the MDFs getting wider with
time, the shift of our overall distribution is approximately $-0.5$
dex in 5 Gyr, which is not compatible with their shift of
approximately $-0.1$ dex in 5 Gyr.

One thing in common for the sets of MDFs presented in Figure \ref{f16}
is that, in spite of a clear mixing, there is a negative metallicity
gradient for the three of them. This point is important because it
tells us the stellar disc can experience significant radial migration
while preserving a clear metallicity gradient. This means that one
should not expect that radial migration is for sure able to erase a
strong metallicity gradient.

\begin{figure*}
\begin{center}
\includegraphics[width=18cm]{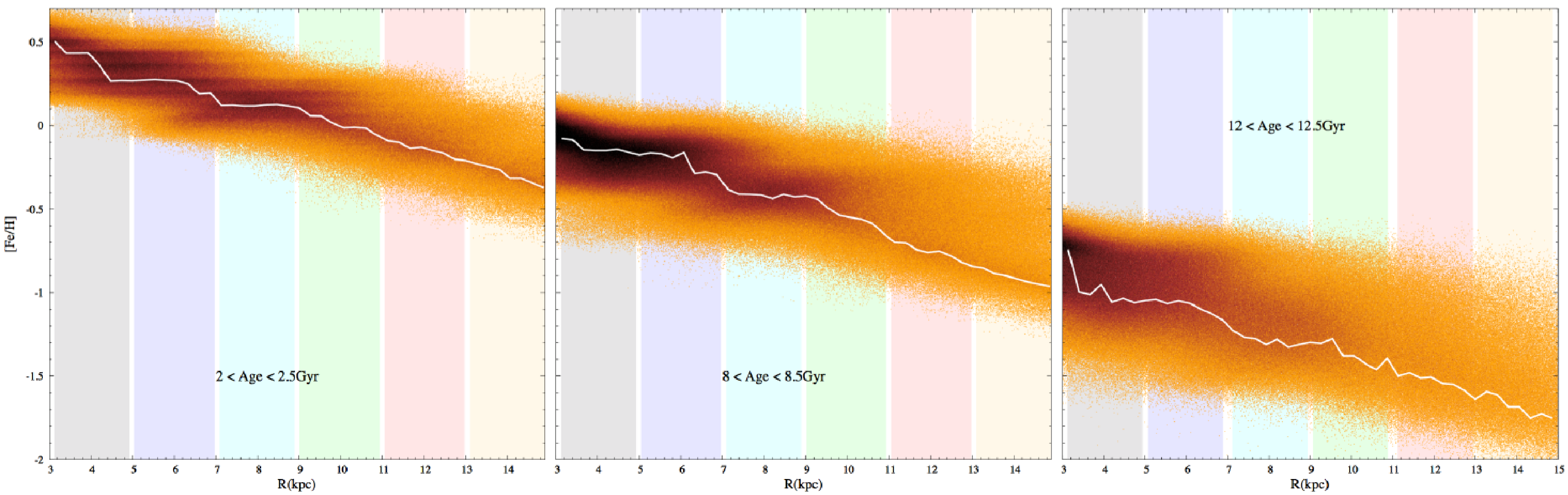}
\end{center}
\caption{Distribution of stars in the plane [Fe/H] vs. Radius for
  populations of three different ages. The color code indicates
  density of stars, while the solid white lines are the mode of the
  [Fe/H] values at every radius.}
\label{f17}
\end{figure*}

This last point is also clear by plotting the same populations in the
[Fe/H] vs. $R$ plane, as shown in Figure \ref{f17}. The stellar
density at every radius shows a global metallicity gradient across the
disc.  Also, the width of the orange bands in the metallicity
distribution is directly correlated with age. It is narrow for the
youngest population, while it is wider for the sample of older stars,
indicating a system more dynamically evolved, and hence, a more mixed
population.

\section{Discussion} 
\label{discussion}

Throughout this work, we have developed some ideas we consider relevant
that we would like to highlight next.

Radial migration can produce an important effect in a galaxy and yet
not be noticed with simple diagnostics such as a single abundance
curve. Indeed, in spite of the fact that radial mixing is a global and
general effect in disc galaxies, erasing the global metallicity
gradient of a given galaxy requires powerful non-axisymmetric features
in the disc; this does not seem to be the case for the MW, as deduced
from a more suitable diagnostic to measure the impact of radial
migration in the MW, like the MDFs. The MDFs seem to be a key
observable to understand the radial migration effect in the MW, which
in turn allows us to understand better the morphology of our
galaxy. With an extensive set of detailed orbital simulations
specifically adjusted to the MW, we conclude that the existence of a
metallicity gradient in a given galaxy, does not rule out the presence
of important radial mixing. Also, radial mixing in the external part
of the Galaxy, is mainly due to the spiral arms and not to the bar;
this means that similar spiral arms in a galaxy, would produce similar
abundance curves and MDFs specifically towards the outskirts of the
galaxy, independently of the existence of a bar. We find that the
corotation resonance is able to produce a local flattening (around it)
in the metallicity global curve (see Figures \ref{f9} and \ref{f17}).

In order to reproduce the Galactic MDFs observed by
\citet{Hayden2015}, our simulations need 22$\%$ flatter Fe/H gradients
during the last 5 Gyr of the evolution than the gradients predicted by
the chemical evolution model (between 8 and 13 Gyr) of
\citet{2011RMxAA..47..139C}. The current Fe/H slope predicted by the
chemical evolution model is $-0.102$ dex kpc$^{-1}$ and its
22$\%$-reduced slope is $-0.080$, for the gaseous component; for the
stellar component, this becomes $-0.063$, which perfectly agrees with
the Fe/H gradient from Cepheids \citep[$-0.062 \pm 0.002$ dex
  kpc$^{-1}$,][]{2011AJ....142..136L} in the 4-16 kpc range (however
this work studies younger stars than the ones observed in the MDF
APOGEE sample). The value of the reduced slope remains constant during
the last 4.5 Gyr of the evolution, particularly in the 6-20 kpc range
\citep[see Fig. 5 by][]{1996RMxAA..32..179C}.

In a chemical evolution model, a flatter Fe/H gradient implies a
flatter O/H gradient by a similar proportion
\citep{1996RMxAA..32..179C}. As mentioned in Section \ref{tagging},
the chemical evolution model was built to reproduce the O/H gradient
in the Galactic disc. The predicted O/H slope is equal to $-0.057$ dex
kpc$^{-1}$ and its 22$\%$-reduced slope is $-0.045$ dex kpc$^{-1}$
which is consistent with the $-0.052 \pm 0.010$ dex kpc$^{-1}$ value
of \citet{2007ApJ...670..457G}, as well as with the O/H slope
($-0.040\pm0.010$ dex kpc$^{-1}$) by \citet{2013MNRAS.433..382E}, when
NGC 2579, an HII region at $\sim 12.4$ kpc, was added to the sample of
\citet{2007ApJ...670..457G}. In the chemical evolution model by
\citet{2011RMxAA..47..139C}, a flatter O/H gradient can be obtained
modifying the accretion rate; mainly, assuming a less intense
inside-out scenario, i.e., taking into account a $\tau(R)$ less
dependent on $R$, which does not affect the agreement with all
chemical properties mentioned in Section \ref{tagging}.

Finally, we would like to emphasize that, although in our first
attempt to fit the MDFs \citep{MartinezMedina2016}, we utilised plain
stellar birth radius as a proxy for metallicity (i.e. without a
recipe for the chemical abundance distribution), we were
able to closely approximate the MDF shapes, qualitatively
speaking. However, it was not possible to fit the absolute magnitudes
nor the width of the MDF; this is only possible with a combination
between both approaches, dynamics and chemistry. Nonetheless,
comparing qualitatively the MDF shapes constructed by using only the
birth radius of the stars, we can see that the main factors that
sculpt the MDFs in the outskirts of the Galaxy are the specific spiral
arms morphology and dynamics. On the other hand, the initial
metallicity gradient plays a fundamental role in the resulting shape
of the inner MDFs, and becomes the main factor, even more important
than the dynamical effect of the bar.

\section{Conclusions} 
\label{conclusions}
With a suitable combination of a detailed observationally motivated
model of the Milky Way galactic potential, that includes spiral arms
and bar, and a careful recipe to approximate the chemical evolution,
we have performed an extensive set of chemodynamical simulations to
study the history of the metallicity distribution function in the
Galaxy.

Even when, in dynamical terms, radial migration does not leave a
kinematical imprint in the stellar disc, the impact of the mechanism
can be discerned from other observables, such as the metallicity
distribution of the stellar disc.

By comparing the MDFs obtained from a set of Galactic models, by
swapping their morphological and dynamical parameters, we show the
sensitivity of the MDF shapes. This means that the shape of the MDF
contains information of the spiral pattern as well as of the stellar
disc dynamical evolution. In this work we provide the characteristics
of the spiral arms that better fit the MDF shapes (Table
\ref{tab:parameters}).

Also, by tuning our simulated MDFs as close as possible to the
observed ones, we introduce a method to set further constrains to
chemical evolution models. This is helpful along the whole disc, but
specially in the central region, where currently models do not provide
information about the chemical content and its evolution.

We have quantified the reach of radial migration in moving stars away
from their birth radius. For example, a considerable number of stars
currently located at the solar circle were born at much smaller
galactocentric distances. By inducing such extended radial excursions,
migration explains how some stars exhibit a chemical content that does
not correspond with the current location of a such star. In this
manner, migration, mainly due to the spiral pattern, models the
metallicity distribution across the disc, imprinting its morphology
and dynamics on the MDFs.

Hundreds of strongly migrating stars that mimic closely the current
solar orbital parameters are found. This shows that the solely study
of radial migration would not be able to accurately constraint the
birth place of the Sun. On the other hand, considering the low
probability that the Sun was formed near its current galactocentric
position, it may be, at least, inaccurate to employ the Sun as
representative of the chemical abundances in the solar neighborhood; a
better way to estimate the birth radius of the Sun, is to combine
chemical and dynamical information. From our chemical model, we obtain
a value of $7.7 \pm 1$ kpc; from our dynamical simulations we obtain
$8.18 \pm 1.38$ kpc. When combining these two determinations, we
obtain that the Sun was born at a galactocentric distance of $7.9 \pm
0.8$ kpc.

We found that a dynamically evolved stellar population can exhibit a
clear metallicity gradient regardless of the presence of important
radial migration and heating in the disc. This means that the presence
of important radial migration does not imply necessarily a substantial
flattening of the metallicity gradient. Therefore, looking for
flattened metallicity gradients is not a trustworthy method to
establish the presence and importance of radial migration in galaxies.

By fitting the separation of the MDFs of the five radial bins
presented by APOGEE, we find a galactic [Fe/H] abundance gradient of
$-0.080$ dex kpc$^{-1}$.

\section*{Acknowledgements}
We would like to acknowledge the anonymous referee for a careful
review and several insightful suggestions. We acknowledge DGTIC-UNAM
for providing HPC resources on the Cluster Supercomputer
Miztli. L.A.M.M and B.P. acknowledge CONACYT Ciencia B\'asica grant
255167. DGAPA-PAPIIT through grants IN-114114 and IG-100115, L.A.M.M
acknowledges DGAPA-PAPIIT through grant IN105916. A.P. acknowledges
DGAPA-PAPIIT through grant IN-109716. L.A.M.M. acknowledges support
from DGAPA-UNAM postdoctoral fellowship.


\end{document}